\PassOptionsToPackage{unicode}{hyperref}
\PassOptionsToPackage{hyphens}{url}
\PassOptionsToPackage{dvipsnames,svgnames,x11names}{xcolor}
\documentclass[
  12pt]{article}

\usepackage{amsmath,amssymb}
\usepackage{iftex}
\ifPDFTeX
  \usepackage[T1]{fontenc}
  \usepackage[utf8]{inputenc}
  \usepackage{textcomp} 
\else 
  \usepackage{unicode-math}
  \defaultfontfeatures{Scale=MatchLowercase}
  \defaultfontfeatures[\rmfamily]{Ligatures=TeX,Scale=1}
\fi
\usepackage{lmodern}
\ifPDFTeX\else  
\fi
\IfFileExists{upquote.sty}{\usepackage{upquote}}{}
\IfFileExists{microtype.sty}{
  \usepackage[]{microtype}
  \UseMicrotypeSet[protrusion]{basicmath} 
}{}
\makeatletter
\@ifundefined{KOMAClassName}{
  \IfFileExists{parskip.sty}{%
    \usepackage{parskip}
  }{
    \setlength{\parindent}{0pt}
    \setlength{\parskip}{6pt plus 2pt minus 1pt}}
}{
  \KOMAoptions{parskip=half}}
\makeatother
\usepackage{xcolor}
\makeatletter
\ifx\paragraph\undefined\else
  \let\oldparagraph\paragraph
  \renewcommand{\paragraph}{
    \@ifstar
      \xxxParagraphStar
      \xxxParagraphNoStar
  }
  \newcommand{\xxxParagraphStar}[1]{\oldparagraph*{#1}\mbox{}}
  \newcommand{\xxxParagraphNoStar}[1]{\oldparagraph{#1}\mbox{}}
\fi
\ifx\subparagraph\undefined\else
  \let\oldsubparagraph\subparagraph
  \renewcommand{\subparagraph}{
    \@ifstar
      \xxxSubParagraphStar
      \xxxSubParagraphNoStar
  }
  \newcommand{\xxxSubParagraphStar}[1]{\oldsubparagraph*{#1}\mbox{}}
  \newcommand{\xxxSubParagraphNoStar}[1]{\oldsubparagraph{#1}\mbox{}}
\fi
\makeatother

\usepackage{longtable,booktabs,array}
\usepackage{calc} 
\usepackage{etoolbox}
\makeatletter
\patchcmd\longtable{\par}{\if@noskipsec\mbox{}\fi\par}{}{}
\makeatother
\IfFileExists{footnotehyper.sty}{\usepackage{footnotehyper}}{\usepackage{footnote}}
\makesavenoteenv{longtable}
\usepackage{graphicx}
\makeatletter
\def\maxwidth{\ifdim\Gin@nat@width>\linewidth\linewidth\else\Gin@nat@width\fi}
\def\maxheight{\ifdim\Gin@nat@height>\textheight\textheight\else\Gin@nat@height\fi}
\makeatother
\setkeys{Gin}{width=\maxwidth,height=\maxheight,keepaspectratio}
\makeatletter
\def\fps@figure{htbp}
\makeatother

\makeatletter
\@ifpackageloaded{caption}{}{\usepackage{caption}}
\AtBeginDocument{%
\ifdefined\contentsname
  \renewcommand*\contentsname{Table of contents}
\else
  \newcommand\contentsname{Table of contents}
\fi
\ifdefined\listfigurename
  \renewcommand*\listfigurename{List of Figures}
\else
  \newcommand\listfigurename{List of Figures}
\fi
\ifdefined\listtablename
  \renewcommand*\listtablename{List of Tables}
\else
  \newcommand\listtablename{List of Tables}
\fi
\ifdefined\figurename
  \renewcommand*\figurename{Figure}
\else
  \newcommand\figurename{Figure}
\fi
\ifdefined\tablename
  \renewcommand*\tablename{Table}
\else
  \newcommand\tablename{Table}
\fi
}
\@ifpackageloaded{float}{}{\usepackage{float}}
\floatstyle{ruled}
\@ifundefined{c@chapter}{\newfloat{codelisting}{h}{lop}}{\newfloat{codelisting}{h}{lop}[chapter]}
\floatname{codelisting}{Listing}

\makeatother
\makeatletter
\@ifpackageloaded{caption}{}{\usepackage{caption}}
\@ifpackageloaded{subcaption}{}{\usepackage{subcaption}}
\makeatother

\ifLuaTeX
  \usepackage{selnolig}  
\fi
\usepackage[]{natbib}
\usepackage{bookmark}

\IfFileExists{xurl.sty}{\usepackage{xurl}}{} 
\hypersetup{
  pdftitle={Title},
  pdfauthor={Author 1; Author 2},
  pdfkeywords={3 to 6 keywords, that do not appear in the title},
  colorlinks=true,
  linkcolor={blue},
  filecolor={Maroon},
  citecolor={Blue},
  urlcolor={Blue},
  pdfcreator={LaTeX via pandoc}}

\newcommand{\anon}{1}

\usepackage{algorithm}
\usepackage{algpseudocode}
\usepackage[normalem]{ulem}
\usepackage{xspace}
\newtheorem{theorem}{Theorem}[section]

\newcommand{\pos}{x}
\newcommand{\velo}{v}
\newcommand{\mo}{p}

\newcommand{\e}{\varepsilon}

\newcommand{\p}{\scalebox{1}[1.4]{$\iota$}}
\newcommand{\Tt}{T}

\newcommand{\ts}{{t^*}}

\newcommand{\Us}{U_\mathrm{sur}}
\newcommand{\Ut}{U_\mathrm{tar}}
\newcommand{\Utf}{U_{\mathrm{tar}, f}}
\newcommand{\Ud}{U_\mathrm{dif}}
\newcommand{\Udf}{U_{\mathrm{dif}, f}}
\newcommand{\Usf}{U_{\mathrm{sur}, f}}

\newcommand{\pr}{\mathbb{P}}

\newcommand{\transpose}{\text{\raisebox{.5ex}{$\intercal$}}}
\newcommand{\diff}{\operatorname{\mathrm{d}}\!{}}

\newcommand{\superH}{{\scalebox{.65}{$\mathrm{H}$}}}
\newcommand{\superB}{{\scalebox{.65}{$\mathrm{B}$}}}
\newcommand{\superP}{{\scalebox{.65}{$\mathrm{P}$}}}

\newcommand{\PDMP}[1][0]{%
	\ifthenelse{\equal{#1}{1}}%
	{\textsc{Pdmp}}%
	{\textsc{pdmp}}%
}

\newcommand{\BPS}[1][0]{%
	\ifthenelse{\equal{#1}{1}}%
	{\textsc{Bps}}%
	{\textsc{bps}}%
}
\newcommand{\HMC}[1][0]{%
	\ifthenelse{\equal{#1}{1}}%
	{\textsc{Hmc}}%
	{\textsc{hmc}}%
}
\newcommand{\ESS}[1][0]{%
	\ifthenelse{\equal{#1}{1}}%
	{\textsc{Ess}}%
	{\textsc{ess}}%
}

\newcommand{\MCMC}[1][0]{%
	\ifthenelse{\equal{#1}{1}}%
	{\textsc{Mcmc}}%
	{\textsc{mcmc}}%
}
\newcommand{\PDMCMC}[1][0]{%
	\ifthenelse{\equal{#1}{1}}%
	{\textsc{Pd-mcmc}}%
	{\textsc{pd-mcmc}}%
}

\newcommand{\HBPS}[1][0]{%
	\ifthenelse{\equal{#1}{1}}%
	{\textsc{Hbps}}%
	{\textsc{hbps}}%
}

\newcommand{\bps}{bouncy particle sampler}
\newcommand{\bhd}{bouncy Hamiltonian dynamics}
\newcommand{\hbps}{Hamiltonian bouncy particle sampler}

\newcommand{\nuts}{no-U-turn algorithm}
\newcommand{\hmc}{Hamiltonian Monte Carlo}
\newcommand{\pdmps}{piecewise-deterministic Markov processes}
\newcommand{\Pdmps}{Piecewise-deterministic Markov processes}

\newcommand{\hzz}{Hamiltonian zig-zag sampler}
\newcommand{\zz}{zig-zag sampler}

\definecolor{emerald}{RGB}{12, 166, 151}
\definecolor{lava}{rgb}{0.81, 0.06, 0.13}

\begin{document}

\def\spacingset#1{\renewcommand{\baselinestretch}%
{#1}\small\normalsize} \spacingset{1}


\if1\anon
{
  \title{\bf MCMC using \textit{bouncy} Hamiltonian dynamics: A unifying framework for Hamiltonian Monte Carlo and piecewise deterministic Markov process samplers}
\author{Andrew Chin
	\\
	and \\
	Akihiko Nishimura \\
	Department of Biostatistics, Bloomberg School of Public Health, \\ Johns Hopkins University}
\maketitle
} \fi

\if0\anon
{
  \bigskip
  \bigskip
  \bigskip
  \begin{center}
    {\LARGE\bf MCMC using \textit{bouncy} Hamiltonian dynamics: A unifying framework for Hamiltonian Monte Carlo and piecewise deterministic Markov process samplers}
\end{center}
  \medskip
} \fi

\bigskip

\vspace*{-.5\baselineskip}
\begin{abstract}
	Piecewise-deterministic Markov process (\PDMP) samplers constitute a state-of-the-art Markov chain Monte Carlo paradigm in Bayesian computation, with examples including the zig-zag and \bps{} (\BPS). 
	Recent work on the zig-zag has indicated its connection to Hamiltonian Monte Carlo (\HMC), a version of the Metropolis algorithm that exploits Hamiltonian dynamics.
	Here we establish that, in fact, the connection between the two paradigms extends far beyond the specific instance.
	The key lies in (1) the fact that any time-reversible deterministic dynamics provides a valid Metropolis proposal and (2) how \PDMP{}s' characteristic velocity changes constitute an alternative to the usual acceptance-rejection.
	We turn this observation into a rigorous framework for constructing rejection-free Metropolis proposals based on \textit{bouncy Hamiltonian dynamics} which simultaneously possess Hamiltonian-like properties and generate discontinuous trajectories similar in appearance to \PDMP{}s.
	When combined with periodic refreshment of the inertia, the dynamics converge strongly to \PDMP{} equivalents in the limit of increasingly frequent refreshment.
	We demonstrate the practical implications of this new framework with a sampler based on a bouncy Hamiltonian dynamics closely related to the \BPS. 
	The resulting sampler exhibits competitive performance on challenging real-data posteriors involving tens of thousands of parameters. 
	As the sampler of choice in modern probabilistic programming languages, \HMC{} plays a critical role in applied Bayesian modeling; by generalizing the paradigm and elucidating its connection to the leading competitor, our framework opens up opportunities for cross-pollination and innovation to further scale Bayesian inference.
\end{abstract}

\noindent%
{\it Keywords:} Bayesian statistics, generalized hybrid Monte Carlo, parameter constraints, event-driven Monte Carlo, probabilistic programming languages
\vfill

\newpage
\spacingset{1.75} 

\section{Introduction}
\label{sec:intro}
Markov Chain Monte Carlo plays a key role in Bayesian inference. 
The Metropolis-Hastings algorithm \citep{metropolis1953equation} provides a general recipe for constructing Markov chains with desired target distributions and has seen widespread use for Bayesian computation.
However, traditional variants of Metropolis-Hastings can struggle in high-dimensional problems due to random walk behavior. 
This has led to increasing popularity of the modern variant known as Hamiltonian Monte Carlo (\HMC) \citep{duane1987hybrid, neal2011mcmc}, especially with its adoption by probabilistic programming languages for applied Bayesian modeling \citep{carpenter2017stan, pymc2016}. 
\HMC[1]{} improves mixing by guiding exploration of the parameter space through Hamiltonian dynamics.
These dynamics utilize the gradient of the target distribution to inform the evolution of an auxiliary momentum variable and thereby the sampler.

Recently, another sampling paradigm has emerged and garnered significant interest in the Bayesian computation community \citep{dunson2020hastings}. 
Built on the theory of \pdmps{} (\PDMP) \citep{davis1984piecewise, fearnhead2018piecewise}, the new paradigm includes the \bps{} (\BPS) \citep{peters2012rejection, bouchard2018bouncy} and \zz{} \citep{bierkens2019zig} as prominent examples. 
These samplers have origins in the non-reversible samplers of \cite{diaconis2000analysis}, who, inspired by the role of momentum in \HMC{}, introduced auxiliary variables that encode senses of direction to discrete-state Markov chains.
These auxiliary variables correspond to velocity variables in \PDMP{}s, which help suppress random walk behavior by promoting persistent motion in one direction at a time.
Instantaneous changes in the velocities occur according to a Poisson process and ensure the correct stationary distribution.

Despite intense research efforts on both algorithms, there has been limited interaction between the two fields.
Previous connections include the work of \citet{nishimura2024zigzag}, who establish a relationship between the \zz{} and a novel variant of \HMC{} using Laplace distributed momentum.
A relationship between \HMC{} and the \BPS{} appears as a high dimensional limit in \citet{deligiannidis2021randomized}, who find that the first coordinate of the \BPS{} process converges to Randomized \HMC{} \citep{bou2017randomized}.
However, this result only concerns a univariate marginal converging as the target dimension tends to infinity.
Another critical distinction from our work is that Randomized \HMC{} is actually an instance of a \PDMP{} and is intrinsically stochastic.
The algorithm gets its name because the differential equation underlying the \PDMP{} admits an interpretation as Hamilton's equations, but it lacks an arguably quintessential feature of the original \HMC{} paradigm and its extensions \citep{fang2014compressible}: the use of a deterministic dynamics as a proposal mechanism.

In this work, we establish a single paradigm unifying the two.
The unification critically relies on our construction of deterministic dynamics that (1) generate trajectories paralleling those of many \PDMP{} samplers in appearance but (2) possess Hamiltonian-like properties allowing their use as a proposal mechanism within the Metropolis algorithm.
We draw inspiration from existing ideas and distill them into two key aspects to construct this novel class of dynamics.
The first idea is the concept of surrogate transition methods \citep{liu2001monte}, which consider the use of generic time-reversible dynamics for proposal generation. 
\HMC[1]{} is a special case in which the dynamics generates rejection-free proposals in the absence of numerical approximation error.
The method is more generally valid under arbitrary Hamiltonian-like dynamics, albeit without the guarantee of high acceptance rates. 
The second idea is that, in \MCMC{} algorithms guided by an auxiliary velocity variable \citep{gustafson1998guided}, an acceptance-rejection step can be replaced by a ``bounce'' event with instantaneous changes in velocity preserving the target distribution. 
A continuous time limit of this idea is used by \citet{peters2012rejection} for motivating the \BPS{}, though the idea has its roots as far back as event driven Monte Carlo of \citet{alder1959studies}.

Our \bhd{} combine the above two ideas by starting with Hamiltonian dynamics corresponding to a surrogate potential energy, which may or may not correspond to the target, and correcting for the surrogate's deviation from the target through a deterministic bounce rule.
We construct this deterministic bounce mechanism through a novel parameter augmentation scheme, introducing an inertia variable that evolves so as to exactly counterbalance discrepancies between the surrogate dynamics' stationary distribution and the target.
Our framework takes any position-velocity dynamics which is time-reversible, volume preserving, and energy conserving, and by bouncing when the inertia runs out, creates a rejection-free proposal mechanism.

These bouncy Hamiltonian dynamics generate trajectories reminiscent of and, in fact, possessing a direct connection to, \PDMP{}s.
By considering a modification in which the inertia variable is periodically refreshed, we show that the bouncy Hamiltonian dynamics converge strongly to \PDMP{} counterparts in the limit of increasingly frequent inertia refreshment. 
The limiting class of processes encompasses the \BPS{} as well as the Boomerang sampler of \citet{bierkens2020boomerang}.
This class of \PDMP{}s is referred to, unfortunately for the nomenclature, as Hamiltonian \PDMCMC{} by \citet{vanetti2017piecewise}.

After establishing the general theory, we turn to demonstrating its practical implications by studying a sampler based on a specific bouncy Hamiltonian dynamics.
We refer to this sampler as the \hbps{} (\HBPS), as its refreshment limit coincides with the \BPS{}.
Theoretically, \HBPS{} dominates the Gaussian random walk Metropolis in sampling efficiency.
Practically, it constitutes an efficient rejection-free sampler for log-concave targets.

We demonstrate the \HBPS{}'s competitive performance on two high-dimensional posteriors from modern real-data applications: a sparse logistic regression for an observational study comparing blood anti-coagulants \citep{tian18lsps} and a phylogenetic probit model applied to \textsc{hiv} data \citep{zhang2021large}. 
Posterior computations for these models rely on Gibbs samplers whose conditional updates include challenging 22,174 dimensional and 11,235 dimensional targets. 
When applied to these conditional updates, our sampler shows superior performance against alternatives and is far easier to tune than the \BPS{}.

We close with discussions of how to extend the practical scope and theory of \bhd{}.
We first introduce a numerical approximation method that enables use of our dynamics when exact solutions for the surrogate dynamics or bounce times are not available.
We then discuss two extensions of bouncy Hamiltonian dynamics: 
one analogous to local \PDMP{} methods \citep{bouchard2018bouncy, vanetti2017piecewise} based on factorization of the target 
and the other based on a novel coordinate-wise scheme applicable also to \PDMP{}s.
Importantly, the latter framework shows the Hamiltonian zig-zag of \cite{nishimura2024zigzag} as a special case of coordinate-wise bouncy Hamiltonian dynamics.

Our bouncy generalization lays the groundwork for novel development in \HMC{}, the algorithm that remains essential in modern Bayesian applications and on which the newest probabilistic programming languages continue to depend \citep{ge2018t, phan2019composable, surjanovic2023pigeons}.
With its appealing properties, \PDMP{}s have been considered a worthy competitor to \HMC{} \citep{corbella_automatic_2022}.
On the other hand, our unifying theory suggests that a significant part of observed differences in their performances is potentially attributable to implementation details, rather than to theoretical sampling efficiencies.
For example, as \cite{corbella_automatic_2022} and \cite{sherlock2022discrete} noted while empirically comparing the two paradigms, inferior performance of \HMC{} under specific situations may be due to the fact that the standard implementation based on Gaussian momentum is sensitive to a target distribution's tail behavior \citep{livingstone2022barker}.
If true, this can be easily remedied by using more robust variants of \HMC{} \citep{livingstone2019kinetic,nishimura2020discontinuous} or our proposed \HBPS{}.
Our unifying theory shows that, more generally, insights and ideas developed in the context of one paradigm may be applicable to the other.
We can advance the field as a whole by cross-pollinating the two paradigms.

\section{Background}
\label{sec:background}

\subsection{\hmc{} and use of time-reversible dynamics to generate Metropolis proposals}
\label{subsec:hmc}
We start this section by reviewing the standard \HMC{} algorithm and proceed to describe how time-reversible, volume preserving dynamics can be used to construct Metropolis proposals more generally.

We denote a target distribution as $\pi(\pos)$, which may be unnormalized, for a parameter of interest $\pos\in \mathbb{R}^d$.
\HMC[1]{} is a version of the Metropolis algorithm which proposes new states through simulation of Hamiltonian dynamics.
To do this, it augments the target with an auxiliary momentum variable $\mo \in \mathbb{R}^d$ and targets the joint distribution $\pi(\pos, \mo) = \pi(\pos) \pi(\mo)$.

Each iteration of \HMC{} follows a two step process.
First, the current momentum value is discarded and a new value $\mo_0$ is drawn from its marginal $\pi(\mo)$. 
Second, combining the new momentum with the current position to form an initial state $(\pos_0, \mo_0)$, Hamiltonian dynamics defined by the following differential equations are simulated for a set time $\Tt$:
\begin{equation}\label{eq:hamildynam}
	\frac{\diff \pos}{\diff t} = \frac{\partial}{\partial \mo} K(\mo) ,\quad \frac{\diff \mo}{\diff t} = -\frac{\partial }{\partial \pos}U(\pos),
\end{equation}
with the kinetic energy $K(\mo) = -\log \pi(\mo)$ and potential energy $U(\pos)$ which, in the standard \HMC{} setting, is taken as $U(\pos) = -\log \pi(\pos)$.
The state of the dynamics at the final time point, $(\pos_{\Tt}, \mo_{\Tt})$, is then used as a proposal and accepted with the standard Metropolis probability $\min\{1, \pi(\pos_T, \mo_T)/\pi(\pos_0, \mo_0)\}$.
Note that the proposal generation step is completely deterministic; randomness in \HMC{} comes only from the momentum refreshment step.

Hamiltonian dynamics exhibits a number of key properties that makes it suitable for generating proposals.
The dynamics is time-reversible and volume preserving, thereby constituting a Metropolis proposal mechanism.
Additionally, the dynamics is energy conserving: the sum of the potential and kinetic energies is constant along its trajectories.
Since $\pi(\pos, \mo) = \exp(-[U(\pos) + K(\mo)])$, the energy conservation implies $\pi(\pos_T, \mo_T) = \pi(\pos_0, \mo_0)$ and makes \HMC{} proposals rejection-free when the dynamics is simulated exactly.
In particular, this property serves as an important inspiration in our construction in Section \ref{sec:general} of \bhd{}; the surrogate dynamics on its own does not conserve the total energy, but the energy conservation is imposed via the additional inertia variable.

\begin{figure}
	\centering
	\begin{minipage}[c]{\textwidth}
		\centering
		\includegraphics[width=0.8\textwidth]{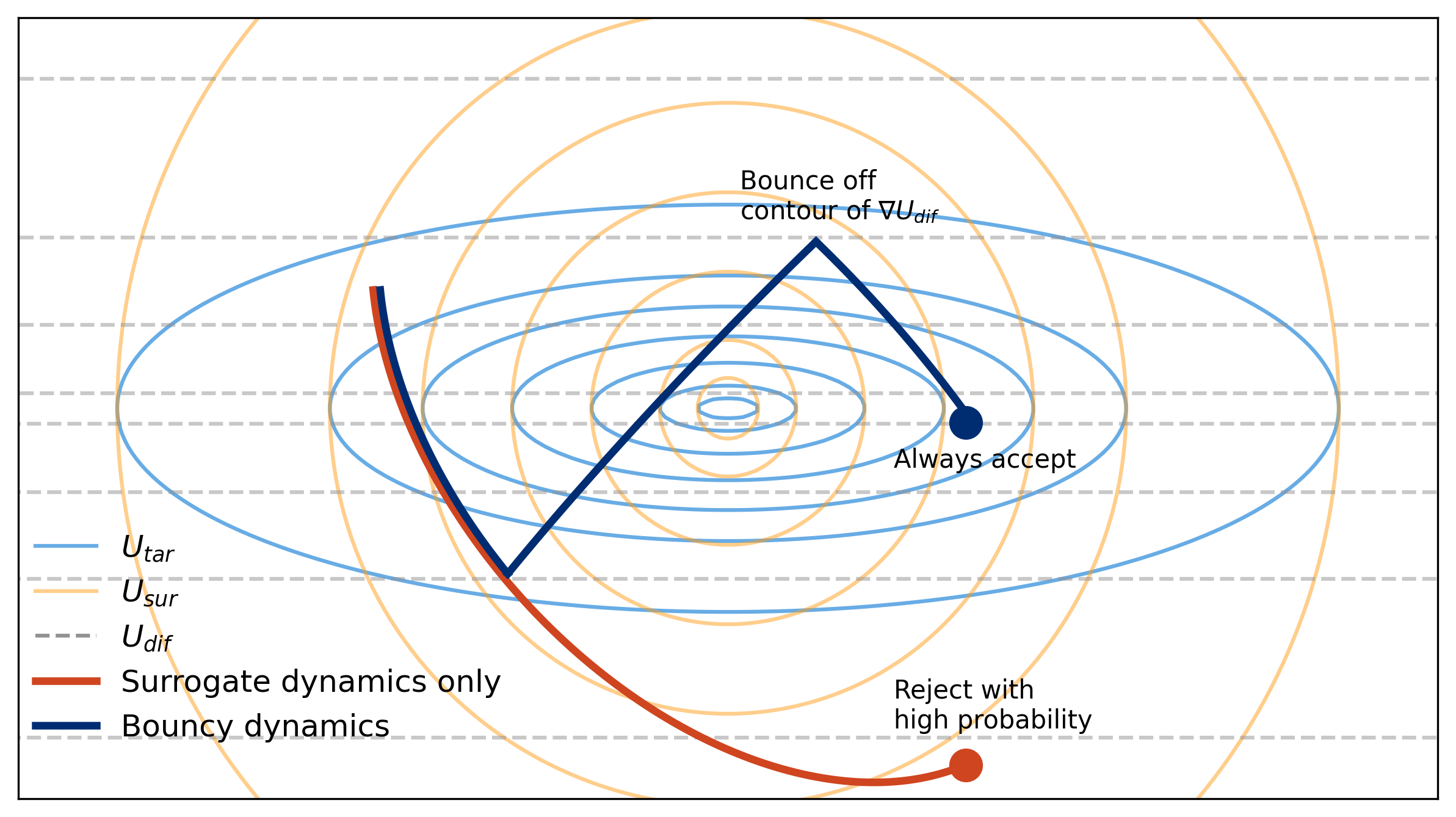}
		\caption{%
			Illustration of proposal generations through surrogate dynamics.
			Here we consider the use of a surrogate $\Us(\pos) = x_1^2/2 + x_2^2/2$ to sample from the target $\Ut(\pos) = - \log \pi(\pos) = x_1^2/2 + 9x_2^2/2$. 
			The red curve shows a trajectory of surrogate dynamics, the orange lines $\Us$'s equipotential contours, and the light blue lines $\Ut$'s contours.
			The grey dotted lines show the contours of $\Ud= \Ut - \Us = 8x_2^2/2$.
			While the red trajectory's end point constitutes a valid Metropolis proposal, it lies in a low probability region of $\Ut$ and is likely to be rejected.
			Our bouncy dynamics (dark blue), to be defined in Section~\ref{sec:general}, yields a trajectory that deterministically reflects against hyperplanes orthogonal to $-\nabla \Ud$ and thereby compensates for the discrepancy between $\Us$ and $\Ut$.
			These deterministic reflections keep the trajectory in high probability regions and ensure its end point to be rejection-free.
			The discontinuous change in velocity occurs deterministically as an integral part of the  bouncy dynamics and, in particular, is distinct from the velocity refreshments commonly used in \HMC{} and \PDMP{} to ensure ergodicity.
		}
		\label{fig:surrogate_vs_bounch}
	\end{minipage}
\end{figure}

While \HMC{} and most of its variants take the potential energy $U$ in Equation \eqref{eq:hamildynam} to coincide with $-\log \pi(\pos)$, Hamiltonian dynamics remain a valid proposal under \textit{any} choice of $U$.
When considering this more general use case, we refer to Hamiltonian dynamics used to generate proposals as \textit{surrogate} dynamics.
To clearly distinguish the potential energy underlying surrogate dynamics and the negative log density of the target, we denote the former as $\Us$, the latter $\Ut$, and their difference $\Ud = \Ut - \Us$.
Since the surrogate dynamics conserves the quantity $\Us(\pos) + K(\mo)$ along the trajectory, the acceptance probability of a generated proposal takes the form
\begin{equation}\label{eq:hmcacceptance}
	\min(1, \exp[-\{\Ud(\pos_T) - \Ud(\pos_0)\} ]).
\end{equation}

\noindent An example of using surrogate dynamics to generate proposals is illustrated in Figure~\ref{fig:surrogate_vs_bounch}.

The potential use of arbitrary Hamiltonian dynamics as a proposal mechanism has not received much attention in the \HMC{} literature, where most research has focused on how to pick the momentum distribution $\pi(\mo)$ \citep{girolami2011riemann,livingstone2019kinetic} and how to numerically approximate the dynamics efficiently \citep{blanes2014numerical, bou2018geometric}.
As we dissect \PDMP{}s in the next section, the surrogate transition perspective helps us view its bounce mechanism as an alternative to an acceptance-rejection procedure.

\subsection{\Pdmps{} as continuous limit of Metropolis moves and delayed-rejection bounces}
\label{subsec:pdmp}
In contrast to deterministic Hamiltonian dynamics, \PDMP{}s are stochastic processes which achieve guided exploration of the original parameter space through an auxiliary velocity variable $\velo$.
\PDMP[1]s target the joint distribution $\pi(\pos) \pi(\velo)$ by combining deterministic dynamics with an inhomogeneous Poisson process that causes random changes in velocity. 

A general class of \PDMP{}s, which subsumes a number of prominent examples, is presented by \citet{vanetti2017piecewise} and is given the name Hamiltonian \PDMCMC{}. 
These \PDMP{}s are so called because they use Hamiltonian dynamics as their deterministic components. 
Different choices of potential energy $U(\pos)$ give rise to different variants, but the kinetic energy is fixed as $K(\velo) = \lVert \velo \rVert ^2/2$, corresponding to the standard Gaussian distribution for $\velo$.
Hamiltonian \PDMCMC{} hence uses the deterministic dynamics
\begin{equation}\label{eq:pdmpdynam}
	\frac{\diff\pos}{\diff t} = \velo_t, \quad \frac{\diff\velo}{\diff t} = - \nabla U(\pos_t).
\end{equation}
\noindent We note that referring to the auxiliary variable as ``velocity'' is appropriate here since the notions of momentum and velocity in Hamiltonian dynamics coincide when assuming the standard Gaussian for the auxiliary variable.
In other words, if we define the auxiliary variable as ``momentum'' $\mo$ and the quantity $\frac{\partial}{\partial \mo} K(\mo)$ providing the time derivative $\diff x/\diff t$ in Equation \eqref{eq:hamildynam} as ``velocity'' $\velo$, then we indeed have $\mo = \velo$ when $K(\mo)= -\log \pi(\mo) = \lVert \mo \rVert^2/2$.

The deterministic component of Hamiltonian \PDMCMC{} plays a role analogous to the surrogate dynamics providing a proposal mechanism as introduced in Section \ref{subsec:hmc}. 
Instead of the acceptance-rejection used by the surrogate transition method, however, Hamiltonian \PDMCMC{} introduces velocity switch events along the trajectory to conserve the target. 
These events occur according to the arrival times of an inhomogeneous Poisson process.
To highlight resemblance to the surrogate transition method, we now reintroduce notation from Section \ref{subsec:hmc}.
We refer to the potential energy $U$ underlying the deterministic component as $\Us$ and set $\Ut = -\log \pi(\pos)$, so $\Ud = \Ut - \Us = -\log \pi(\pos) - U$.
With this notation, the Poisson rate for velocity switches in Hamiltonian \PDMCMC{} is given by
\begin{equation}\label{eq:poisrate}
	\lambda(t) = \lambda(\pos_t, \velo_t) =  \{\velo_t^\transpose \nabla \Ud(\pos_t)\}^+ = \left \{\left.\frac{\diff}{\diff s} \Ud(\pos_t )\right|_{s=t} \right \}^+
\end{equation}
and can be seen as a function of the discrepancy between the surrogate and target.

By a property of Poisson processes, the velocity switch event time can be simulated by drawing an exponential random variable $\e$ with unit rate and then solving for
\begin{equation}\label{eq:bpsbounce}
	t^*_\superB = \inf_{t>0} \left [\e = \int_0^t \{\velo_t^\transpose \nabla \Ud(\pos_t)\}^+ \diff s \right].
\end{equation}

\noindent At time $t^*_\superB$, Hamiltonian \PDMCMC{} undergoes an instantaneous change in the velocity $\velo \gets R_\pos(\velo)$ in the form of an elastic bounce off the hyperplane orthogonal to $\nabla \Ud$, where the reflection function $R_\pos(\velo)$ is defined by
\begin{equation}\label{eq:reflect}
	R_\pos(\velo) = \velo - 2 \frac{\velo^\transpose \nabla \Ud(\pos)}{\lVert \nabla \Ud(\pos) \rVert^2}	\nabla \Ud(\pos).
\end{equation}
Hamiltonian \PDMCMC{} as prescribed samples from the given target as can be verified straightforwardly, after the fact, using Kolmogorov's equations for Markov jump processes \citep{fearnhead2018piecewise}.
However, it is apposite to ask: where does the algorithm come from in the first place?
In tracing the historical development of \PDMP{}s, we find the algorithm to emerge from two key ideas. 
The first is to guide a Markov chain's movement via an auxiliary variable to promote persistent motion \citep{diaconis2000analysis}, and this idea's connection to \PDMP{}s is most clearly seen in the guided random walk  of \citet{gustafson1998guided}. The second is the idea of ``event-driven'' moves as alternatives to acceptance-rejection \citep{alder1959studies, peters2012rejection} which, in statistical language, can be interpreted as a delayed-rejection scheme \citep{mira2001metropolis, sherlock2022discrete}.

The guided random walk uses an auxiliary direction variable $v$ and makes moves in the position-velocity space.
We present a multivariate version of the algorithm which is an extension of the original work by \citet{gustafson1998guided} implicitly used by \citet{sherlock2022discrete}.
The algorithm draws the auxiliary variable $v\in \mathbb{R}^d$ from a standard multivariate Gaussian and makes a move of the form:
\begin{equation}\label{eq:grw}
	(\pos_{\Delta t}^{n+1}, \velo_{\Delta t}^{n+1}) = 
	\begin{cases}
		(\pos_{\Delta t}^{n} + \Delta t \velo_{\Delta t}^{n}, \velo_{\Delta t}^{n})  & \text{with prob.\ }  \displaystyle \min\left \{1, \frac{\pi(\pos_{\Delta t}^{n} + \Delta t \velo_{\Delta t}^{n})}{\pi(\pos_{\Delta t}^{n})}\right\}  \\
		(\pos_{\Delta t}^{n}, -\velo_{\Delta t}^{n})  & \text{otherwise}.
	\end{cases}
\end{equation}
Crucially, when a proposal is accepted, $v$ is maintained so the next proposal is again in the same direction; when proposals are rejected, $v$ is negated so the next proposal is in the reverse direction.

The proposal $\pos_n + \Delta t \velo_n$ can be viewed as simulating Gaussian momentum based Hamiltonian dynamics for time $\Delta t$ on a flat potential with initial position and velocity $(\pos_n, \velo_n)$.
This perspective allows us to generalize the guided random walk to use Hamiltonian dynamics with arbitrary potential $\Us$.
Let $\Phi(t, \pos, \velo): \mathbb{R}^{2d+1} \to \mathbb{R}^d$ denote the \textit{solution operator} of this dynamics, which takes an initial state $(\pos, \velo)$, and returns the state $(\pos_t, \velo_t)$ obtained as a result of simulating the dynamics for time $t$.
Our generalized guided random walk proposes a Hamiltonian move  $(\pos_{\Delta t}', \velo_{\Delta t}') = \Phi(\Delta t, \pos, \velo)$ and accepts it with probability 
\[
\min\left(1, \exp\!\left[ -\{ \Ud(\pos_{\Delta t}') - \Ud (\pos)\} \right] \right)
\]
or negates the velocity otherwise.
This results in a Markov chain which starts backtracking the trajectory of Hamiltonian dynamics upon rejection.

In the limit  $\Delta t \to 0$, this process converges to a continuous time Markov chain where the velocity flips at the first arrival time of a Poisson process with rate given in Equation \eqref{eq:poisrate}.
This can be seen by applying second order Taylor expansions to the acceptance probability:
\begin{align*}
	\min\left(1, \exp\!\left[ -\{ \Ud(\pos_{\Delta t}') - \Ud (\pos)\} \right] \right)
	&= \exp\!\left[ -\{ \Ud(\pos_{\Delta t}') - \Ud (\pos)\}^+ \right] \\
	&= \exp\!\left[ -\{ \Delta t (\velo^\transpose \nabla \Ud(\pos)) + O(\Delta t^2) \}^+ \right] \\
	&= 1- \Delta t\{\velo^\transpose \nabla \Ud(\pos)\}^+ + O(\Delta t^2).
\end{align*}
The rejection probability for each move is thus approximately $\Delta t \{\velo^\transpose \nabla \Ud(\pos)\}^+$, and there are $\lfloor \Tt / \Delta t \rfloor$ attempted moves when simulating the dynamics for duration $\Tt$.

To avoid the backtracking behavior, which restricts exploration to one dimension, we introduce the second key idea of event-driven bounces as an alternative to simply negating the velocity.
More precisely, we incorporate the reflective bounce of Equation~\eqref{eq:reflect} into the guided Markov chain using a delayed rejection scheme.
Upon rejection, we can generate a new proposal $(\pos''_{\Delta t}, \velo''_{\Delta t}) = \Phi\{\Delta t,\pos'_{\Delta t}, R_{\pos'_{\Delta t}}(\velo'_{\Delta t})\}$ to be accepted with probability \citep{mira2001metropolis, sherlock2022discrete}
\[\max\left(1, \frac{1- \exp[-\{\Ud(\pos'_{\Delta t}) - \Ud (\pos''_{\Delta t})\}^+]}{1- \exp[- \{\Ud(\pos'_{\Delta t}) - \Ud (\pos)\}^+]} \frac{\exp\{-\Ut(x''_{\Delta t})\}}{\exp\{-\Ut(x)\}}\right). \]
In the guided walk's continuous limit $\Delta t \to 0$, the above acceptance probability converges to $R_\pos(-\velo)^\transpose \nabla \Ud (\pos)/\velo^\transpose \nabla \Ud (\pos) = 1$ and results in a bounce event that occurs according to the Poisson rate \eqref{eq:poisrate}.

The above construction shows Hamiltonian \PDMCMC{} as a limit of a discrete-time Markov chain guided by surrogate Hamiltonian dynamics combined with delayed-rejection bounces.
The derivation of the \BPS{} as a continuous limit in \citet{peters2012rejection} and \citet{sherlock2022discrete} can be viewed as a special case in which the surrogate Hamiltonian dynamics corresponds to a trivial one with $\Us = 0$.

\section{Bouncy Hamiltonian Dynamics}\label{sec:general}

\subsection{Constructing rejection-free dynamics by synthesizing surrogate dynamics with event-driven bounces}
The surrogate method discussed in Section \ref{subsec:hmc} generates proposals using surrogate dynamics with potential energy $\Us$ and corrects for the discrepancy between $\Us$ and $\Ut$ via acceptance-rejection. The acceptance probability in Equation \eqref{eq:hmcacceptance} degrades as the discrepancy grows, leading to a large proportion of rejections and inefficient exploration. Here we show that acceptance-rejection can be replaced by deterministic bounces in the trajectory.
This resulting dynamics piecewise follow the surrogate dynamics in between deterministic bounces and generate rejection-free proposals.

Our construction of \bhd{} starts with surrogate dynamics whose trajectory $(\pos_t, \velo_t)$ is given by the solution to Equation $\eqref{eq:hamildynam}$ with surrogate potential $\Us(\pos)$ and kinetic energy $K(\velo)=\lVert \velo \rVert^2/2$.
In introducing deterministic bounces, we take inspiration from \citet{nishimura2024zigzag}, who find that Hamiltonian dynamics with Laplace distributed momentum yields trajectories akin to the zig-zag process. 
Crucially, they find this Hamiltonian dynamics to undergo an instantaneous velocity switch at the moment when momentum runs out along a coordinate.
Emulating this behavior, we introduce an inertia variable $\p \geq 0$ that is exponentially distributed with unit rate, and define the bounce time as the moment when the inertia runs out. 
The dynamics operate in $\mathbb{R}^{2d+1}$ to target the joint distribution $\pi(\pos, \velo, \p) = \pi(\pos) \pi(\velo)\pi(\p)$, whose negative log density is $\Ut(\pos) + K(\velo) + \p$.
This inertia will play the role of a pseudo-momentum, and, additionally, evolve so as to conserve the joint density $\pi(\pos, \velo, \p)$.

From an initial state $(\pos_0, \velo_0, \p_0)$, we let the position-velocity component $(\pos_t, \velo_t)$ evolve according to the surrogate dynamics.
In the meantime, we define the evolution of the inertia component as 
\begin{equation}\label{eq:generaldynamics}
	\p_t = \p_0 -  \int_0^t \velo_s^\transpose \nabla \Ud(\pos_s)\diff s= \p_0 + \Ud(\pos_0) - \Ud(\pos_t),
\end{equation}
\noindent where we denote $\Ud = \Ut - \Us$ as before.
With this specification, the inertia runs out at time
\begin{equation}\label{eq:bouncetime}
	\ts = \inf_{t>0} \left\{\p_0 = \int_0^t \velo_s^\transpose \nabla \Ud(\pos_s)\diff s  = \Ud(\pos_t) -\Ud(\pos_0) \right\}.
\end{equation} 
At time $\ts$, a reflection occurs against $\nabla \Ud$ in the manner of Equation \eqref{eq:reflect}, emulating the bounce of the \PDMP.
The process for simulating the dynamics for duration $\Tt$ is summarized in Algorithm \ref{alg:hbpd}, where $\Phi(t, \pos, \velo)$ as before denotes the solution operator of the surrogate dynamics.  
Figure~\ref{fig:surrogate_vs_bounch} illustrates a trajectory using only the surrogate proposals compared to our bouncy dynamics.

\spacingset{1.15} 
\noindent\begin{minipage}[b]{0.50\textwidth}
	\begin{algorithm}[H]
		\centering
		\caption{\texttt{BouncyHamiltonian
				Dynamics}($\Tt, \pos, \velo, \p$)}\label{alg:hbpd}
		\begin{algorithmic}
			\State $\tau \gets 0$\Comment{current time traveled}
			\While{ $\tau < \Tt$}
			\State $\ts = \inf_{t>0} \{\p = \int_0^t \velo_{\tau+s}^\transpose \nabla \Ud(\pos_{\tau+s})\diff s \}$
			\If{ $\tau + \ts > \Tt$}
			\State $\pos_\Tt, \velo_\Tt \gets \Phi(\Tt-\tau, \pos, \velo)$
			\State $\p_\Tt \gets \p_\tau - \int_0^{\Tt -\tau} \velo_{\tau +s}^\transpose \nabla \Ud(\pos_{\tau+s})\diff s $ 
			\State \Return  $\pos_\Tt, \velo_\Tt, \p_\Tt$
			\Else
			\State $\pos_{\tau+\ts}, \velo_{\tau+\ts} \gets \Phi(\ts, \pos_\tau, \velo_\tau)$
			\State $\velo_{\tau+\ts} \gets R_{\pos_{\tau+\ts}}(\velo_{\tau+\ts})$
			\State $\p_{\tau+\ts} \gets 0$
			\State $\tau \gets \tau + \ts$
			\EndIf
			\EndWhile
		\end{algorithmic}
	\end{algorithm}
\end{minipage}
\hfill
\begin{minipage}[b]{0.48\textwidth}
	\begin{algorithm}[H]
		\caption{Bouncy Hamiltonian sampler for $n$ iterations.}\label{alg:hbpssampler}
		\begin{algorithmic}
			\Require $\pos, T, n$
			\For{$i=1,\dots, n$}
			\State Draw $\velo \sim N(0, I)$
			\State Draw $\p \sim Exp(1)$
			\State $\pos, \velo, \p \gets $ \texttt{BouncyHamiltonian \\ \qquad Dynamics}$(\Tt, \pos, \velo, \p)$ \Comment{Algorithm \ref{alg:hbpd}}
			\State Store $\pos$ as sample
			\EndFor
		\end{algorithmic}
	\end{algorithm}
	\vspace{89pt}
\end{minipage}
\spacingset{1.75} 
The \bhd{} as constructed maintains the surrogate dynamics' ability to generate valid proposals and, further, makes proposals rejection-free by exactly compensating for the discrepancy between the surrogate and target potential:

\sloppy
\begin{theorem}\label{thm:general}
	Assume that $\Us$ is twice continuously differentiable and the set $\{(\pos, \p): \nabla \Ud(\pos) = 0, \  \p=0\} \cup \{(\pos, \velo, \p): \velo^\transpose\nabla \Ud(\pos) = 0, \  \p=0\}$  consists of smooth manifolds of dimension at most $d-1$.
	Then the corresponding \bhd{} is well-defined on $\mathbb{R}^{2d+1}$ away from a set of measure 0 and is time-reversible, volume preserving, and conserves the augmented energy $\Ut(\pos) + K(\velo) + \p  = -\log\pi(\pos, \velo, \p)$.
\end{theorem}
\fussy

\noindent The proof is in Supplement A. 
Particularly remarkable is the dynamics' volume preservation property;
this is automatic for classical Hamiltonian dynamics because of their symplecticity \citep{bou2018geometric}, 
but the bouncy dynamics defined on odd-numbered dimensions cannot possibly be symplectic and its volume preservation has to be established from first principles.

The properties established by Theorem \ref{thm:general} together imply our bouncy dynamics constitute a valid Metropolis proposal mechanism; 
Algorithm~\ref{alg:hbpssampler} summarizes the steps for the resulting sampler.
We also note that the \bhd{} can be straightforwardly extended to accommodate parameter constraints by elastically reflecting off boundaries in the same manner as in standard Hamiltonian dynamics and \PDMP{}s \citep{neal2011mcmc, bierkens2018pdmp_on_restricted_domain}.

We close this section by discussing notable special cases of bouncy Hamiltonian dynamics.
When setting $\Us = \Ut$, no bounces occur because $\Ud = 0$ and we recover classical, smooth Hamiltonian dynamics.
The choice $\Us = \|\pos\|^2/2$ yields dynamics whose trajectories parallel those of the Boomerang sampler \citep{bierkens2020boomerang}; 
this bouncy dynamics can be used to construct an analogue of the Boomerang sampler, though we do not pursue this possibility here.
Instead, for the rest of this article, we focus on the case $\Us = 0$.
This yields a bouncy Hamiltonian analogue of the \BPS{}'s piecewise linear dynamics with constant velocity, which we call \HBPS{} dynamics.
We study this dynamics and resulting sampler further in Section \ref{sec:hbps}.

\subsection{Piecewise-deterministic Markov processes as a limit}\label{subsec:limit}
The trajectories of the \bhd{} and its \PDMP{} counterpart have much in common: both follow the same deterministic dynamics between events and undergo bounces in an identical manner.
On the other hand, the two dynamics differ in how event times are determined: one follows a deterministic process dictated by the inertia variable, while the other a Poisson schedule dictated by exponential random variables. 

We now present a theoretical result that quantifies a relationship between the two dynamics. 
To this end, we consider modifying \bhd{} by adding resampling of inertia from its marginal at regular intervals of size $\Delta t$.
This refreshment of inertia introduces randomness in otherwise deterministic bounce times. 
As we increase the frequency of this refreshment by letting $\Delta t \to 0$, the next bounce time becomes effectively independent of the past evolution, making the dynamics Markovian.
In fact, in this limit, the dynamics converges in its position and velocity components to its \PDMP{} counterpart.
To state the result more formally, let $D[0, \infty)$ denote the space of right-continuous-with-left-limit functions from $[0, \infty) \to \mathbb{R}^d \times \mathbb{R}^d$ with the Skorokhod topology \citep{billingsley1999convergence, ethier2009markov}. 
We also make a common assumption in the \HMC{} literature to control the dynamics' behavior in the tails \citep{livingstone2019geometric}, which roughly says the density proportional to $\exp(-\Us)$ has tails no lighter than exponential.
We then have the following result:

\begin{theorem}[Weak Convergence]\label{thm:weak}
	Assume $\Us$ and $\Ut$ are twice continuously differentiable and $\lim \sup_{\lVert \pos \rVert \to \infty} \frac{\lVert \nabla \Us(\pos) \rVert}{\lVert \pos \rVert} < \infty$.
	Consider a sequence of \bhd{}, indexed by $\Delta t >0$, with a common initial position and velocity $(\pos_0, \velo_0)$ but with the added inertia refreshment at every $\Delta t$ interval. 
	Let $(\pos_{\Delta t}^\superH, \velo_{\Delta t}^\superH)$ denote the position and velocity components of these dynamics.
	As $\Delta t \to 0$, the position-velocity dynamics $(\pos_{\Delta t}^\superH, \velo_{\Delta t}^\superH)$ converges weakly in $D[0, \infty)$ to the corresponding \PDMP{} with the same initial condition and inter-bounce dynamics governed by the same surrogate Hamiltonian dynamics.
\end{theorem}
In fact, the above result is a consequence of a stronger convergence result below.

\begin{theorem}[Strong Convergence]\label{thm:strong}
	Under the same assumptions as in Theorem \ref{thm:weak}, the position-velocity dynamics $(\pos_{\Delta t}^\superH, \velo_{\Delta t}^\superH)$ converges strongly to the same corresponding \PDMP{} limit in $D[0,T]$ for any $T>0$.
	More precisely, we can construct a sequence of \PDMP{}s $(\pos_{\Delta t}^\superP, \velo_{\Delta t}^\superP)$, all of which have the same distribution as the limiting \PDMP, coupled to the \bhd{} so that
	\[ \lim_{\Delta t \to 0 }\pr[\rho_T\{(\pos_{\Delta t}^\superH, \velo_{\Delta t}^\superH), (\pos_{\Delta t}^\superP, \velo_{\Delta t}^\superP)\} > \epsilon] = 0 
	\ \text{ for any } \, \epsilon > 0, \]
	where $\rho_T$ denotes the Skhorokod metric on $[0,T]$.
\end{theorem}

\noindent Since the strong convergence holds for any $T>0$, Theorem \ref{thm:strong} implies the weak convergence result of Theorem \ref{thm:weak} \citep{billingsley1999convergence}.
The proof of Theorem \ref{thm:strong} is in Supplement~B. 

The theorem shows that \bhd{} not only generalize classical Hamiltonian dynamics, but also recover \PDMP{}s in the limit, thereby unifying the \HMC{} and \PDMP{} paradigms.
One implication is that the theoretical performance of the two paradigms may be closely related.
On the other hand, as we demonstrate in Section \ref{sec:app}, their deterministic and stochastic nature result in significant differences in their practical performance.

\section{The \hbps}\label{sec:hbps}

\subsection{Bouncy Hamiltonian dynamics underlying the sampler}\label{subsec:hbpd}
\begin{figure}
	\begin{minipage}[c]{0.54\textwidth}
		\centering
		\includegraphics[width=0.8\textwidth]{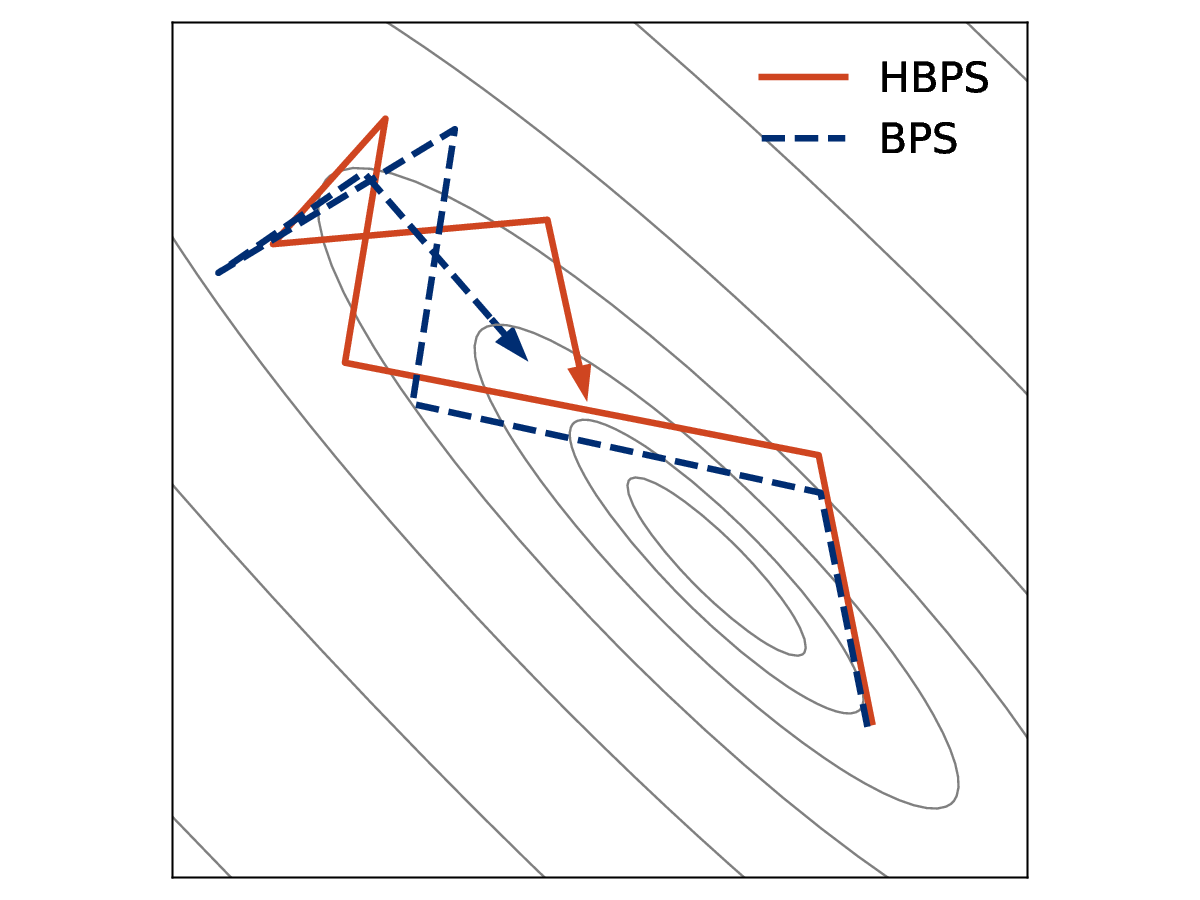}
	\end{minipage}
	\hspace*{-.03\textwidth}
	\begin{minipage}[c]{0.44\textwidth}
		\caption{Trajectories of the \HBPS{} dynamics (solid red) and the \BPS{}'s dynamics (dashed blue) from the same initial position and velocity on a correlated bivariate Gaussian target.}
		\label{fig:trajectories}
	\end{minipage}
\end{figure}

The \BPS{} is an instance of a Hamiltonian \PDMCMC{} algorithm as described in Section \ref{subsec:pdmp} with the surrogate dynamics corresponding to a constant potential.
Since the dynamics is only affected up to additive constants, we assume without loss of generality that $\Us=0$ and $\Ud = \Ut$.
The differential equation \eqref{eq:pdmpdynam} governing the deterministic part then becomes 
\[\frac{\diff \pos}{\diff t} = \frac{\partial}{\partial \velo} \frac{\lVert \velo \rVert ^2}{2} = \velo, \quad \frac{\diff \velo}{\diff t} = -\frac{\partial}{\partial \pos} 0 = 0,  \]
From a given initial position and velocity $(\pos_0, \velo_0)$, therefore, the dynamics evolves as
\[
\pos_t = \pos_0 + t\velo_0  ,\quad \velo_t = \velo_0
\]
until velocity switch events that occur according to the rate
\begin{equation}\label{eq:bpsrate}
	\lambda(t) =  \{\velo_0^\transpose \nabla \Ut(\pos_0+t\velo_0)\}^+.
\end{equation}
We now replace the Poisson bounce events of the \BPS{} with our deterministic inertia-driven process.
This yields \bhd{} whose evolution, following Equation \eqref{eq:generaldynamics} and \eqref{eq:bouncetime}, takes the following form.
Prior to a bounce event, the dynamics evolves as 
\begin{equation}\label{eq:hbpsevolve}
	\begin{aligned}
		\pos_t &= \pos_0 + t\velo_0,\quad \velo_t = \velo_0\\
		\p_t &= \p_0 -  \int_0^t \velo_0^\transpose \nabla \Ut(\pos_0+s\velo_0)\diff s  = \p_0 + \Ut(\pos_0) -\Ut(\pos_0 +t\velo_0).
	\end{aligned}
\end{equation}
\noindent The bounce event occurs when the inertia runs out at time $\ts$, given by
\begin{equation}\label{eq:hbpsbounce}
	\ts = \inf_{t>0} \left \{\p_0 = \int_0^t \velo_0^\transpose \nabla \Ut(\pos_0+s\velo_0)\diff s  = \Ut(\pos_0 +t\velo_0) - \Ut(\pos_0) \right\}.	
\end{equation}
\noindent At time $\ts$,  the velocity is reflected in exactly the same manner as the \BPS{}:
\begin{align*}
	\pos^* &= \pos_0 + \ts \velo_0 ,\quad \velo^* = R_{\pos^*}(\velo_0),\quad	\p^* = 0.
\end{align*}
\noindent With this new state, the next event time is computed and the process is repeated.
Figure~\ref{fig:trajectories} compares an example trajectory to the \BPS{}'s.

Thanks to Theorem \ref{thm:general}, the above dynamics can be used as a valid proposal mechanism.
We term the resulting version of the bouncy Hamiltonian sampler as the \textit{\hbps} (\HBPS) and the underlying deterministic dynamics as \textit{\HBPS{} dynamics}.

We show in Supplement C that, despite all its Hamiltonian-like properties, \HBPS{}  dynamics does not admit any interpretation as a bona fide Hamiltonian dynamics within existing theories.
This fact shows that our bouncy Hamiltonian dynamics is a fundamentally new proposal generation mechanism that expands the tool kit for Bayesian computation.

\subsection{Efficient implementation on log-concave targets}\label{subsec:sampler}
\HBPS[1]{} dynamics is of particular interest as it permits exact simulation on log-concave targets; 
since its trajectory is linear in the position component in between bounces, solving for the bounce times boils down to convex optimization.
This enables a rejection-free sampler which is provably more efficient than random walk Metropolis:

\begin{theorem}\label{thm:mh_dominance}
	On a log-concave target, a random walk Metropolis sampler with an isotropic Gaussian proposal is always dominated in asymptotic efficiency by the \HBPS{} with travel time equal to the proposal standard deviation.
	More precisely, for any function $f$ square integrable with respect to $\pi(\cdot)$, the ergodic average $n^{-1} \sum_{i = 1}^{n} f(\pos^{(i)})$ under the random walk sampler has a larger asymptotic variance than that under the \HBPS{}.
\end{theorem}
The superior efficiency of \HBPS{} relates to the fact that its bounce mechanism is a deterministic, continuous-time analogue of the delayed-rejection scheme, which delivers an efficiency gain guaranteed via the Peskun-Tierney ordering \citep{mira2001metropolis}.
The proof is Supplement~D.

The dynamics' deterministic and time-reversible nature also makes it possible to take advantage of the \nuts{} of \citet{hoffman2014no} to minimize tuning.
This algorithm only requires a base unit of step size as an input and multiplicatively adapts the total travel time  for efficient exploration while preserving detailed balance.
The base step size can be chosen either heuristically or adaptively \citep{nishimura2024zigzag}.
The combination of \HBPS{} dynamics with no-U-turn results in a rejection-free sampler with minimal user input.
This sampler in particular excels in Gibbs sampling schemes where the conditional posteriors change at each iteration and may require varying travel times for optimal exploration of the conditional.

Simulating the \HBPS{} is more computationally efficient than simulating the \BPS{} on log-concave targets.
Solving for the \BPS{}'s event time $t^*_\superB$ in Equation \eqref{eq:bpsbounce}, which contains a positive-part operator in the integrand, requires a minimization first to find the time $t^*_{min}$ at which the directional derivative turns positive \citep{bouchard2018bouncy}:
\[
t^*_{min} = \inf_{t>0} \{ \velo_0^\transpose \nabla \Ut(\pos_0+t\velo_0) \geq 0  \}.
\]
\noindent Then, an additional root finding procedure is required to find $t^*_\superB$:
\[
t^*_\superB = \inf_{t>t^*_{min}} \left \{\e = \int_{t^*_{min}}^t \velo_0^\transpose \nabla \Ut(\pos_0+s\velo_0) \diff s = \Ut(\pos_0+t\velo_0) - \Ut(\pos_0 + t^*_{min} \velo_0) \right\}.
\]
\noindent On the other hand, computing bounces times for the \HBPS{} requires only a single root finding algorithm to solve Equation \eqref{eq:hbpsbounce}, bypassing the need for the minimization.
This significantly reduces the number of required operations relative to the \BPS{}.

For our applications, we solve for the roots using the Newton--Raphson algorithm for both samplers.
Minima for the \BPS{} are found using the \textsc{bfgs} algorithm, which we find to be faster for our applications than Newton--Raphson.

Constraints on the position parameter space can be handled in the same way as the \BPS{} by elastically bouncing off of the boundaries \citep{bouchard2018bouncy}.
The required computation of boundary event times is trivial as both samplers have constant velocities.

\section{Applications}\label{sec:app}
\subsection{Setup}\label{subsec:appsetup}
We compare the \HBPS{} against other algorithms on two challenging real-data applications.  
In each example, the joint posterior distribution is difficult to sample from directly, but the conditional distributions are more tractable.
We thus employ a Gibbs scheme where the computational bottleneck is a high dimensional log-concave distribution. We apply the \HBPS{} and alternative algorithms to this conditional update and compare variations of the Gibbs samplers differing only in this aspect.
Performance is measured using effective sample size (\ESS) calculated by the method of \citet{plummer2006coda} and normalized by computation time.
We measure \ESS{} across the parameters of interest and then use the minimum value achieved, representing how effectively the algorithm sampled the most difficult parameter.
Final results are averaged over five runs with different seeds.
We present \ESS{} per time relative to the performance of the \BPS{} to facilitate comparisons, with \BPS{} taking the value of 1 and higher values indicating higher \ESS{} per time.

The comparisons feature three algorithms with increasing numbers of tuning parameters: the \HBPS{} with \nuts{}, the \HBPS{} with manually-tuned travel time, and the \BPS{} with manually-tuned travel time and refresh rate. 
The refresh rate is necessitated by the known risk of \BPS{}'s reducible behavior when using the event rate in Equation \eqref{eq:bpsrate} on its own \citep{bouchard2018bouncy}.
The sampler is thus typically implemented with additional velocity refreshment by introducing a refresh rate $\lambda_{ref}$ and using the combined event rate $\tilde \lambda(t) = \lambda(t) +\lambda_{ref}$. 
At the first arrival time $\ts$ of this combined rate, the process refreshes the velocity with probability $\lambda_{ref}/\tilde \lambda(\ts)$ or bounces otherwise.

The base step size for the \nuts{} is chosen as suggested in \citet{nishimura2024zigzag} by estimating the covariance matrix, computing the maximum eigenvalue $\kappa_{max}$, and then using $0.1\sqrt{\kappa_{max}}$.  
No further tuning for the \nuts{} is done to treat it as an option with minimal need for user input. 
For tuning of the \HBPS{}, we start with the average travel time $\Tt$ selected by the \nuts{} as our starting point.
Then we search a grid of travel times of $\Tt \pm j\Tt/4$, for $j=0,1,2,3$.

We run simulations on the Joint High Performance Computing Exchange cluster at Johns Hopkins, allocating a single \textsc{amd epyc 7713 cpu} core and 32 gigabytes of memory for each chain.
Code for Section~\ref{subsec:logreg} is available at \url{https://tinyurl.com/59vrru58} and code for Section~\ref{subsec:phylo} is available at \url{https://tinyurl.com/4r2rk5pf}.

\subsection{Bayesian sparse logistic regression: conditionally log-concave target}\label{subsec:logreg}
The first posterior distribution arises from an observational study comparing risk of gastrointestinal bleeding from two alternative blood anticoagulants, dabigatran and warfarin.
The study is based on the new-user cohort design and includes of 72,489 patients and 22,174 covariates extracted from the Merative MarketScan Medicare Supplemental and Coordination of Benefits Database.
As part of the average treatment effect estimation, propensity scores \citep{rosenbaum1983central} are estimated with a Bayesian sparse logistic regression model, the required posterior computation for which we compare our samplers.

The probability of taking dabigatran is modeled via logistic regression:
\[
\textrm{logit}\{p_i(\beta)\} = \textrm{logit}\{\pr(i\textrm{-th patient on dabigatran})\} = x_i^\transpose \beta.
\]
The regression coefficients $\beta_j$ are given a bridge prior \citep{polson2013bayesian} to induce sparsity in the posterior estimates. 
Under the global-local representation, commonly used to facilitate the posterior computation, the prior is parametrized with global and local scale $\tau$ and $\lambda_j$ as:
\[
\beta_j \mid \tau, \lambda_j \sim N(0, \tau^2 \lambda_j^2), \quad \tau \sim \pi_{\textrm{global}}(\cdot), \quad \lambda_j \sim \pi_{\textrm{local}}(\cdot).
\]
For interested readers, a full description of the data and model can be found in \citet{nishimura2022prior}. 

We compare variations of the Gibbs samplers differing only in their conditional updates of the regression coefficients. 
Directly sampling from the conditional distribution of $\beta$ is expensive due to ill-conditioning, so all samplers draw from the preconditioned scale with $\tilde \beta_j = \beta_j/(\tau \lambda_j)$ \citep{nishimura2022prior}, which has the following log-concave density:
\[
\tilde \beta \mid y, X, \tau, \lambda \propto \left [\prod_i p_i(\tilde \beta)^{y_i}\{1-p_i(\tilde \beta)\}^{1-y_i}\right ]\exp(-\tilde \beta^\transpose\tilde\beta/2).
\]
This conditional update is traditionally dealt with using the P\'{o}lya--Gamma augmentation of \citet{polson2013bayesian}, conditioning on the additional parameter $\omega$ so that the distribution $\beta \mid y, X, \omega, \tau, \lambda$ becomes Gaussian.
We run this augmented Gibbs sampler as a baseline comparison.
The \BPS{} and its bouncy Hamiltonian counterpart forego the augmentation and sample directly from the log-concave distribution.

The samplers are started from the same initial position at stationarity, determined by running 4,000 iterations of the P\'{o}lya--Gamma augmented sampler and taking the final sample. 
With the tuning strategy described in Section \ref{subsec:appsetup}, the base step size for the no-U-turn \HBPS{} was derived to be 0.1, and optimal travel time for the manually-tuned version to be 1.5.
The \BPS{} also requires selecting a travel time and, additionally, a refresh rate. 
We conduct a grid search over both parameters.
For the travel time, we start with 1.5 as chosen for the \HBPS{} and adjust it by increments of $\pm 0.5$.
For the refresh rate, we start with 1 as suggested in \citet{deligiannidis2021randomized} and adjust it by increments of $\pm0.2$.
This approach finds the optimal refreshment rate to be 0.2, the lower boundary of the initial search grid. 
We thus refine our search grid on the lower end so that optimal parameters are within the boundaries of the new grid.
Specifically, we add 0.1, 0.05, 0.01, and 0.005 to the new search grid and confirm a refresh rate of 0.005 not to be the optimal choice.
For the \BPS{}, a travel time of 1.5 is optimal along with a refresh rate of 0.01, with performance being much more sensitive to travel time choice than refresh rate. 

\spacingset{1.15} 
\begin{table}
	\caption{Minimum \ESS{} per hour across all regression coefficients for the sparse logistic model.
		Both raw \ESS{} values and those relative to the \BPS's are provided. \label{tab:1}}
	\vspace{-0.2\baselineskip}
	\begin{center}
		\begin{tabular}{lcc}
			& Relative \ESS{} per time & Raw \ESS{} per hour \\
			P\'{o}lya--Gamma& 0.36 & 0.16\\
			\BPS{} & 1 & 0.45 \\
			\HBPS{} with no-U-turn  & 1.29 & 0.58\\	
			\HBPS{} with manual tuning & 4.02 & 1.81\\	
		\end{tabular}	
	\end{center}
\end{table}
\spacingset{1.75} 

The results are presented in Table \ref{tab:1}. 
All samplers achieved at least 60 effective samples along each dimension.
Due to the scale of the problem, each run took approximately a week to complete.
Over 14,000 hours of total \textsc{cpu} time were required to complete the tuning procedure and generate the final results.

\begin{figure}[h]
	\centering
	\begin{minipage}[c]{\textwidth}
		\centering
		\includegraphics[width=0.8\textwidth]{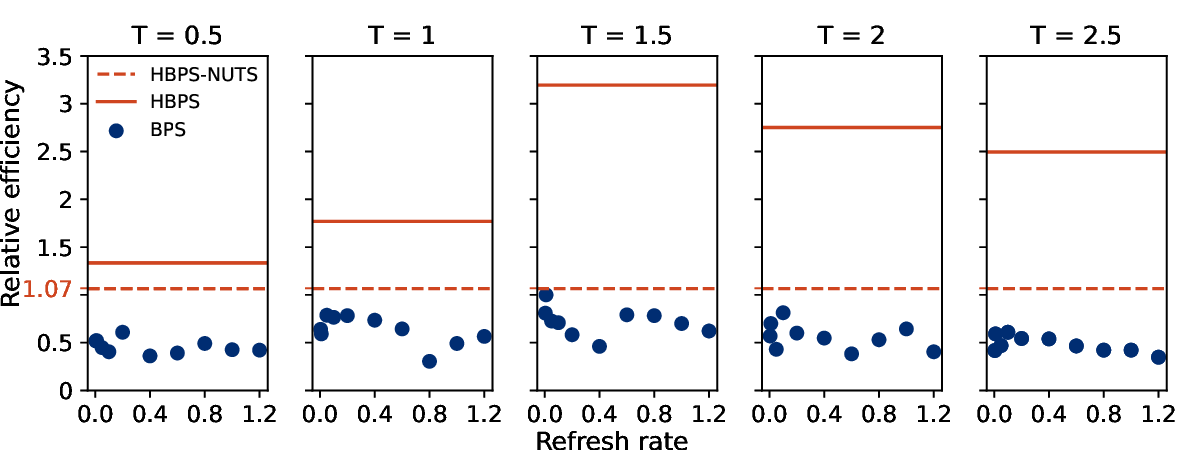}
		\caption{%
			Comparison, based on the sparse logistic model, of the no-U-turn (red dashed) and manually-tuned (red solid) \HBPS{} against the \BPS{} (blue points) under the five best travel time parameters and the grid of refresh rates. 
			The presented efficiency values are relative to the best-performing \BPS{}.
			This optimally-tuned \BPS{} identified here are re-run five times for the final benchmarking result as shown in Table~\ref{tab:1};
			having been estimated from the separate runs, the relative efficiency of \HBPS{} reported there differs slightly from those reported here.
		}
		\label{fig:bpstune}
	\end{minipage}
\end{figure}

In our simulations, the manually-tuned \HBPS{} is the best performing sampler by a factor of four over the \BPS.
The no-U-turn version is less efficient due its paths requiring more computation time, but it remains superior to the \BPS{} under any tuning parameter.
The competitive performance and minimal tuning requirement makes this an attractive option for practical purposes. 
Figure~\ref{fig:bpstune} shows the \BPS{} performance under the extensive range of the tuning parameters tried.
We see that the \BPS{}'s performance is sensitive to its tuning parameters and, if not for our laborious tuning effort, would have lagged behind the \HBPS{}'s by an even larger margin.

The P\'{o}lya--Gamma augmented sampler is the slowest performing sampler overall.
More generally, the P\'{o}lya--Gamma sampler is known to suffer from poor mixing when dealing with imbalanced binomial outcomes \citep{johndrow2019mcmc}. 
\HBPS[1]{}'s ability to forego the data augmentation and directly target the log-concave conditional is, therefore, likely to provide an even greater advantage for binomial outcomes.

\subsection{Phylogenetic probit model: conditionally truncated Gaussian and splitting for joint update}\label{subsec:phylo}
Our second target comes from a model that quantifies the biological covariation among binary traits for the \textsc{hiv} virus while accounting for spurious correlations from shared evolutionary history.
Of scientific interest is identifying which mutations are associated with immune response and viral virulance.
To investigate this, \citet{zhang2021large} employ a Bayesian phylogenetic probit model.
We provide a brief summary of the data and model below, and refer readers to \citet{zhang2021large} for further details.

The data consist of $n=535$ \textsc{hiv} viruses, each with $m=21$ traits, collected from patients in Botswana and South Africa \citep{payne2014impact}. 
The model uses continuous latent biological traits $x \in \mathbb{R}^{m\times n}$ which follow a phylogenetic Brownian diffusion parametrized by a phylogenetic tree $\mathcal{T}$ and diffusion covariance $\Gamma \in \mathbb{R}^{m\times m}$.
The observed binary outcomes $y = \textrm{sign}(x) \in \mathbb{R}^{m\times n}$ indicate the presence of traits in each virus.
The primary parameter of interest is the covariance matrix $\Gamma$, which represents biological covariation independent of phylogenetic history among the $m$ binary traits.
The across-trait correlation and partial correlation matrices give insight into potential causal pathways from mutations to disease security and infectiousness of viruses.

The inference relies on a Gibbs scheme applied to the joint posterior $(x,\Gamma, \mathcal{T})$, whose bottleneck is updating from the conditional $x \mid \Gamma, \mathcal{T}, y$.
As in the application of the previous section, we vary the choice of sampler for this conditional update.
This conditional follows a $535\times 21 = 11,235$ dimensional truncated normal distribution of the form:
\[
x \mid \Gamma, \mathcal{T}, y  \sim N\{\mu(\Gamma, \mathcal{T}), \Sigma(\Gamma, \mathcal{T})\}\,  \text{ truncated to }\,  \{\text{sign}(x)= y\}.
\]
Special features of this posterior make the \HBPS{} an attractive choice of sampler.
The covariance matrix $\Sigma(\Gamma, \mathcal{T})$ depends on the other parameter values and hence changes during each scan of the Gibbs sampler.
This makes exact sampling with \HMC{} \citep{pakman2014exact}, which requires inverting a 11,235 dimensional matrix at each iteration, prohibitively expensive.
On the other hand, the dynamic programming trick of \citet{zhang2021large} allows us to compute matrix-vector multiplications by the precision matrix in $O(np^2)$ time instead of the usual $O(n^2p^2)$.
Both \BPS{} and \HBPS{} rely only on this operation for their implementations.
Another feature is the high correlation between the two parameters $x$ and $\Gamma$ which leads to slower mixing when updating them conditionally \citep{liu2001monte}.
The deterministic and reversible nature of the \HBPS{} dynamics allows for a joint update via an operator splitting scheme \citep{nishimura2020discontinuous, shahbaba2014split}.
The splitting scheme approximates the \textit{joint dynamics} on $(x, \Gamma)$ for time $\Delta t$ via a scheme similar to Section 2.2.2 of \citet{zhang2022accelerating}: we first simulate the dynamics of $\Gamma$ given the initial value of $x$ for a half step $\Delta t/2$, then simulate the dynamics  of $x$ given the updated $\Gamma$ for a full step $\Delta t$, and finally simulate $\Gamma$ given the final position of $x$ for a half step $\Delta t/2$.

\citet{zhang2022accelerating} update $x$ conditionally on $\Gamma$ using the \BPS{}, which we replicate as a baseline using the same parameters.
For comparison, we run two Gibbs samplers utilizing the \HBPS{}.
The first simply updates $x$ with the \HBPS{} and a tuned travel time of 1.
The second uses the joint update of the $\Gamma$ and $x$ with operator splitting, with the number of steps taken decided by the \nuts{}.
For all the samplers and both conditional and joint updates, sampling of $\Gamma$ relies on the Gaussian momentum based Hamiltonian dynamics with leapfrog approximation.
Ten percent of the samples are discarded as burn-in.

We measure minimum \ESS{} across all unique elements of the correlation and partial correlation matrices, ensuring all samplers achieve a minimum \ESS{} of 100.
Final results are normalized by computation time, averaged over five runs with different seeds, and are presented in Table \ref{tab:2}.
Like the sparse logistic regression example in Section \ref{subsec:logreg}, the scale of the data necessitates multiple days of computation for each sampler, with the \BPS{} achieving an average raw \ESS{} per hour of~3.9.

\spacingset{1.15} 
\begin{table}
	\caption{%
		Minimum \ESS{} per hour across all unique elements of correlation and partial correlation matrices for the phylogenetic model. 
		Both raw \ESS{} values and those relative to \BPS{}'s are provided.
	\label{tab:2}}
	\vspace{-0.2\baselineskip}
	\begin{center}
		\begin{tabular}{>{\raggedright}p{4cm} >{\centering}p{1cm} >{\centering}p{1cm} >{\centering}p{.5cm} >{\centering}p{1cm} >{\centering}p{1cm} >{\centering\arraybackslash}p{.5cm}}
			& \multicolumn{6}{c}{\hspace*{-2ex}Raw (relative) \ESS{} per hour} \\
			& \multicolumn{2}{c}{Correlation}  & \multicolumn{4}{c}{Partial correlation}  \\
			\BPS{} & 9.23 & (1) & & 3.92 & (1) & \\
			\HBPS{}& 12.20 & (1.32) & & 3.55 & (0.91) &\\
			\HBPS{} with splitting& 11.82 & (1.28) & & 11.01 & (2.81) & \\				
		\end{tabular}
	\end{center}
\end{table}
\spacingset{1.75} 

For the correlation parameters, all samplers exhibit similar performance.
However, differences are evident for the partial correlation parameters, which \citet{zhang2022accelerating} consider more difficult to sample from.
While the \HBPS{} performs similarly to the \BPS{} when deploying the same conditional update approach, it achieves a 2.8 times speedup when taking advantage of the opportunity to jointly update the parameters via operator splitting.

We note that the data subsampling approach popular in the \PDMP{} literature is not applicable here since the phylogenetic model grows in dimension as the number of observations grows, so there is no redundancy in the data.

\section{Discussion}\label{sec:disc}
In this article, we have presented a framework that unifies two of the most important paradigms in Bayesian computation. 
Our \bhd{} use deterministic bounces to both generalize Hamiltonian dynamics and connect it to \PDMP{}s through an inertia refreshment limit.
We now present two techniques to further generalize \bhd{} and broaden its practical applicability.

\subsection{Numerical approximation of \bhd{}}
In general, surrogate dynamics and/or bounce times cannot be solved for exactly.
While this article demonstrated bouncy Hamiltonian dynamics on examples admitting exact simulations, we now present a numerical approximation scheme based on the splitting technique \citep{strang1968construction, mclachlan2002splitting} to extend the algorithm's scope to more general surrogates and targets.
While splitting is widely used for numerically solving differential equations, our scheme is unique in its ability to approximate dynamics with discontinuous bounces, and is partially inspired by the scheme of \citet{bertazzi2023splitting} for \PDMP{}s.
Outlined in Algorithm \ref{alg:general}, our scheme can be constructed from either an exact solution operator $\Phi$ of the surrogate dynamics or an approximate one $\hat \Phi$ based on, for example, the leapfrog integrator.
Algorithm \ref{alg:general} is time-reversible and volume preserving whenever $\Phi$ or $\hat \Phi$ is time-reversible and volume preserving, and hence generates valid Metropolis proposals.

Our scheme starts by taking a half step of the surrogate dynamics in the position-velocity space.
It then updates the inertia by approximating the formula \eqref{eq:generaldynamics} through a midpoint approximation 
\[\Ud(\pos_{\Delta t}) - \Ud(\pos_0) = \int_0^{\Delta t} \velo_s^\transpose\nabla \Ud(\pos_s) \diff s \approx \Delta t \velo_{\Delta t/2}^\transpose \nabla \Ud(\pos_{\Delta t/2}), \] 
or reflects the velocity if the trajectory does not have sufficient inertia.
This is followed by the remaining half step in position and velocity. 
When using this approximate bouncy dynamics, generated proposals are no longer rejection-free and hence the bouncy Hamiltonian sampler (Algorithm \ref{alg:hbpssampler}) must be modified to include a Metropolis acceptance-rejection step.

\spacingset{1.15}
\begin{algorithm}[H]
	\centering
	\caption{Midpoint integrator with stepsize $\Delta t$ for bouncy Hamiltonian dynamics}\label{alg:general}
	
	\begin{algorithmic}
		\State $\pos_{\Delta t/2}, \velo_{\Delta t/2} \gets \Phi(\Delta t/2, \pos_0, \velo_0)$ or $\hat \Phi(\Delta t/2, \pos_0, \velo_0)$
		\State $\p' \gets \p_0 - \Delta t  \velo_{\Delta t/2}^\transpose\nabla \Ud(\pos_{\Delta t/2})$
		\If{$\p' >0$}
		\State $\p_{\Delta t} \gets \p'$
		\Else
		\State $\p_{\Delta t} \gets \p_0$
		\State $\velo_{\Delta/2} \gets R_{\pos_{\Delta t/2}}(\velo_{\Delta t/2})$
		\EndIf
		\State $\pos_{\Delta t}, \velo_{\Delta t} \gets \Phi(\Delta t/2, \pos_{\Delta t/2}, \velo_{\Delta t/2})$ or $\hat \Phi(\Delta t/2, \pos_{\Delta t/2}, \velo_{\Delta t/2})$
	\end{algorithmic}
\end{algorithm}
\spacingset{1.75}

\subsection{Local and coordinate-wise \bhd{}}
Our \bhd{} can be further generalized in two complementary ways.
The first is akin to local methods described in \citet{bouchard2018bouncy} and \citet{vanetti2017piecewise}.
These methods are useful in applications such as factor graph models when the target density can be represented as a product
$
\pi(x) \propto \prod_{f =1}^F\gamma_f(x),
$
where $\gamma_f$ is referred to as a \textit{factor} and depends only on a subset of the coordinates of $x$ indexed by $N_f \subseteq \{1,\dots, d\}$. 
In this case, the target potential energy becomes a sum
\[
\Ut(\pos) = \sum_{f =1}^F \Utf(\pos),
\]

\noindent where $\Utf= -\log \gamma_f$ has the property that $\partial \Utf(\pos) /\partial \pos_k=0$ for all $k \not \in N_f$. 
A local \PDMP{} assumes a surrogate with the same factorization $\Us = \sum_f \Usf$ and assigns to each factor its own event rate $\lambda_f = [\velo^\transpose \nabla \Udf]^{+}$. When the factor $f$ experiences an event, the bounce occurs against $\nabla \Udf$ and affects the subset $\{ v_k \}_{k \in N_f}$ of the velocity according to the restricted reflection function:
\[
R_\pos(\velo, f) = \velo - 2 \frac{\velo^\transpose \nabla \Udf(\pos)}{\lVert \nabla \Udf(\pos) \rVert^2}	\nabla \Udf(\pos). 
\]
To adapt this localization idea to \bhd{}, we associate to each factor an inertia $\p_{f,0} \sim Exp(1)$ instead of a Poisson process.
These inertias evolve and run out according to local analogues of Equations \eqref{eq:generaldynamics} and \eqref{eq:bouncetime}:
\[
\p_{f,t} = \p_{f,0} -  \int_0^t \velo_s^\transpose \nabla \Udf(\pos_s)\diff s, \quad
t^*_f = \inf_{t>0} \left \{\p_{f,0} = \int_0^t \velo_s^\transpose \nabla \Udf(\pos_s)\diff s \right\}.	
\]

\noindent The bounce occurs at time $\min_f t^*_f$, following the same restricted reflection function as in the local \PDMP.
The process is then repeated with this new velocity.

The second generalization of \bhd{} is obtained by partitioning the coordinates of the parameter space, associating an inertia to each part, and bouncing only on the subset of coordinates whose inertia runs out.
Denote a partition $\mathcal{J}$ such that $\bigcup_{J\in \mathcal{J}} J = \{1, \dots, d\}$.
For each $J \in \mathcal{J}$, denote by $\pos_J, \velo_J \in \mathbb{R}^{|J|}$ the corresponding position and velocity coordinates. 
Then associate an inertia $\p_J \in \mathbb{R}$ to each part $J$ and define its evolution as
\[
\p_{J,t} = \p_{J,0} -  \int_0^t \velo^\transpose_{J,s}\partial_J \Ud(\pos_{s})\diff s,
\]
where $\partial_J$ represents the vector of partial derivatives associated to part $J$.
When $\p_{J,t}$ runs out, $\velo_J$ is reflected off the hyperplane orthogonal to $\partial_J \Ud(\pos_s)$, as in Equation~\eqref{eq:reflect} but with $\velo_J$ and $\partial_J \Ud$ in place of $\velo$ and $\nabla\Ud$.
This coordinate-wise scheme can be combined with the local scheme by partitioning the coordinates associated with each factor.

Both generalizations retain the same properties as outlined in Theorem~\ref{thm:general}, with proofs almost identical to the original case, and thus generate valid Metropolis proposals.
Analogues of Theorems \ref{thm:weak} and \ref{thm:strong} also hold: when combined with the increasingly frequent refreshments of inertias, the dynamics converge to their corresponding \PDMP{}s in the limit.

The coordinate-wise extension is of particular interest as it allows us to subsume the earlier connection between \HMC{} and \PDMP{}s under our present framework. 
Specifically, when partitioning the coordinates into singleton dimensions, the coordinate-wise bouncy dynamics under a constant surrogate potential coincides with the \hzz{} of \citet{nishimura2024zigzag} and converges to the \PDMP{} zig-zag sampler in the limit.
This fact also reflects that the coordinate-wise scheme is applicable to \PDMP{}s in an analogue manner;
applying the singleton coordinate partition to the \bps{}, for example, gives rise to the \zz{}. 
New classes of \PDMP{}s can also be constructed by considering block partitions, providing further potential opportunities.

%
%
%
%
%
%
%
%
%

\spacingset{1.15}
\bibliography{hbps_ref.bib}

\end{document}


\def\spacingset#1{\renewcommand{\baselinestretch}%
{#1}\small\normalsize} \spacingset{1}


\if1\anon
{
  \title{\bf Supplement for ``MCMC using \textit{bouncy} Hamiltonian dynamics: A unifying framework for Hamiltonian Monte Carlo and piecewise deterministic Markov process samplers"}
\author{Andrew Chin
	\\
	and \\
	Akihiko Nishimura \\
	Department of Biostatistics, Bloomberg School of Public Health, \\ Johns Hopkins University}
\maketitle
} \fi

\if0\anon
{
  \bigskip
  \bigskip
  \bigskip
  \begin{center}
    {\LARGE\bf  Supplement for ``MCMC using \textit{bouncy} Hamiltonian dynamics: A unifying framework for Hamiltonian Monte Carlo and piecewise deterministic Markov process samplers"}
\end{center}
  \medskip
} \fi

\spacingset{1.75} 

\appendix

\section{Proof of Theorem \ref{main-thm:general}}\label{supp-pf:general}

We first prove the well-definedness, time-reversibility, and conservation of the augmented energy.
We then separately treat the volume preservation as its proof is more involved.

\begin{proof}(\textit{of well-definedness, time-reversibility, and conservation of the augmented energy})
	\strut\linebreak
	Our construction in Section~\ref{main-sec:general} completely characterizes the dynamics as long as its trajectories avoid, at the moment of bounce $\p =0$, the following two problematic situations. 
	First, if the inertia were to run out at a point where $\nabla \Ud(\pos) =0$, the velocity after the bounce 
	\[
	\velo - 2 \frac{\velo^\transpose \nabla \Ud(\pos)}{\lVert \nabla \Ud(\pos) \rVert^2}	\nabla \Ud(\pos)\]
	would be indeterminate.
	Second, if the inertia were to run out when $\velo^\transpose \nabla \Ud(\pos) =0$, then no actual change in velocity occurs, and this glancing bounce introduces potential ambiguity in what can happen at the next moment.
	In other words, the dynamics are well-defined as long as its trajectories avoid the set
	\[
	\{(\pos, \p): \nabla \Ud(\pos) = 0, \  \p=0\} \cup \{(\pos, \velo, \p): \velo^\transpose\nabla \Ud(\pos) = 0, \  \p=0\}.
	\]
	Following an argument nearly identical to \citet{nishimura2024zigzag}, we can show that there is a measure zero set away from which all trajectories avoid the problematic set for all $t>0$. 
	We omit the proof for brevity and refer interested readers to \citet{nishimura2024zigzag}.
	
	We turn to the question of time-reversibility.
	Let $\Phi(\Tt, \pos_0, \velo_0, \p_0)$ denote the solution operator for \bhd{} that returns the state $(\pos_\Tt, \velo_\Tt, \p_\Tt)$ at time $\Tt$ from the initial condition $(\pos_0, \velo_0, \p_0)$.
	In this setting, time-reversibility is equivalent to the condition $(\pos_0, - \velo_0, \p_0) = \Phi(\Tt, \pos_T, -\velo_T, \p_T)$. 
	That is, flipping the velocity and then simulating the dynamics forward in time recover the initial state up to another velocity flip.
	This property follows from the fact that the underlying position-velocity dynamics is assumed to be reversible and the inertia evolution and the reflection process are trivially reversible.
	
	Conservation of the augmented energy is also straightforward to show.
	Since $\pi(\velo)$ is symmetric in all directions, bounce events do not change the joint density.
	Hence it suffices to show that the density remains constant in between bounces.
	This can be shown via the following equality for the joint negative log density, where the right hand side is clearly constant by the energy conservation property of the surrogate dynamics:
	\[
	\Ut(\pos_t)  + K(\velo_t) + \p_t = \Us(\pos_t)  + K(\velo_t) + \p_0 + \Ud(\pos_0). 
	\] 
\end{proof}

\begin{proof}(\textit{of volume preservation})
	\begingroup
	\everymath{\displaystyle}
	Volume-preservation means that the Jacobian of the mapping $\Phi_\Tt(\pos_0, \velo_0, \p_0)  \mapsto (\posT, \vT, \pT)$ has determinant of magnitude one, which we now prove.
	We can assume without loss of generality that, by dividing the interval into smaller pieces as needed, at most one bounce occurs in the interval $[0, \Tt]$ since a composition of volume preserving maps is still volume preserving. 
	
	In case no bounce occurs in $[0,T]$, the determinant of the $2d+1$ dimensional Jacobian $D\Phi_\Tt$ will be given as follows, where $\0$ represents zero matrices of appropriate dimensions: 
	\begin{align*}
		\det D\Phi_\Tt = 
		\det \begin{bmatrix}
			\frac{\partial \posT}{\partial \pos_0} & \frac{\partial \posT}{\partial \velo_0} & \0 \\
			\frac{\partial \vT}{\partial \pos_0} & \frac{\partial \vT}{\partial \velo_0} & \0 \\
			\frac{\partial \pT}{\partial \pos_0} & \frac{\partial \pT}{\partial \velo_0} & 1 \\
		\end{bmatrix}
		=	\det \begin{bmatrix}
			\frac{\partial \posT}{\partial \pos_0} & \frac{\partial \posT}{\partial \velo_0}  \\
			\frac{\partial \vT}{\partial \pos_0} & \frac{\partial \vT}{\partial \velo_0}  \\
		\end{bmatrix} 
		=1 \text{ or } {-1}.
	\end{align*}
	The last equality is due to volume preservation of the surrogate dynamics.
	
	It remains to establish volume preservation when one bounce occurs at time $\ts \in [0,T]$.
	Let $\phi(t, \pos, \velo, \p): \mathbb{R}^{2d+2} \to \mathbb{R}^{2d+1}$ denote the solution operator of the \bhd{} \textit{without any bounces},
	which takes an initial state $(\pos, \velo, \p)$ and duration $t$ and returns $(\pos_t, \velo_t, \p_t)$, the state of the dynamics after time $t$.
	That is, $\phi(t, \pos, \velo, \p)$ represents the dynamics that continues to evolve according to Equation~\eqref{main-eq:hbpsevolve} regardless of whether it runs out of inertia.
	Setting $z_0 = (\pos_0, \velo_0, \p_0)$, the full dynamics $\Phi_\Tt$ \textit{with a bounce} at time $\ts = \ts(z_0)$ can then be represented as 
	\[
	\Phi_\Tt(z_0) = \phi[
	\Tt - \ts(z_0), R \circ \phi\{\ts(z_0),  z_0\} 
	].
	\]
	We now show that $\Phi_\Tt$ has Jacobian with absolute determinant one.
	For brevity we denote the state instantaneously before and after the bounce by $\zsm =  \phi(\ts,  z_0)$ and $\zsp =  R \circ \phi(\ts,  z_0)$.
	To facilitate computation, we let
	\[
	f(t, z) = \phi\{\Tt-t, R \circ \phi(t,  z)\}
	\]
	so that $\Phi_\Tt(z_0) = f\{\ts(z_0),  z_0\}$.
	We can now differentiate $\Phi_\Tt(z_0)$ with respect to $z_0$ via the chain rule to obtain:
	\begin{equation}
		\label{eq:rank_one_perturbation_representation}
		\begin{aligned}
			D\Phi_\Tt 
			= \left. \frac{\partial f}{\partial t} \right |_{\ts, z_0}\frac{\partial \ts}{\partial z_0} + \left. \frac{\partial f}{\partial z} \right |_{\ts, z_0} 
			= uw^\transpose +M,
		\end{aligned}
	\end{equation}	
	where the last expression can be viewed as a rank-1 pertubation of a $(2d+1) \times (2d+1)$ matrix with
	\[w^\transpose =\frac{\partial \ts}{\partial z_0} \in \mathbb{R}^{1\times(2d+1)}\] 
	\[u = \left. \frac{\partial f}{\partial t} \right |_{\ts, z_0}
	= \left.-\frac{\partial \phi}{\partial t}\right|_{\Tt-\ts,  \zsp} + \left.\frac{\partial \phi}{\partial z}\right|_{\Tt-\ts,  \zsp} 	
	\left.\frac{\partial R}{\partial z} \right |_{\zsm} 
	\left.\frac{\partial \phi}{\partial t} \right |_{\ts, z_0}  \in \mathbb{R}^{2d+1} \]
	\[M = \left. \frac{\partial f}{\partial z} \right |_{\ts, z_0} 
	= \left.\frac{\partial \phi}{\partial z}\right|_{\Tt-\ts,  \zsp} 
	\left.\frac{\partial R}{\partial z}\right |_{\zsm} 
	\left.\frac{\partial \phi}{\partial z}\right |_{\ts, z_0} \in \mathbb{R}^{(2d+1)\times(2d+1)}.\]
	
	By applying the matrix determinant lemma to the expression \eqref{eq:rank_one_perturbation_representation}, we have that
	\[
	\det D\Phi_\Tt = (1+ w^\transpose M^{-1} u)\det M.
	\]
	We have shown that \bhd{} is volume preserving in the absence of bounce of events and hence $\phi$ is also volume preserving.
	We thus have $\det \frac{\partial \phi}{\partial z} = 1$.
	Combining this with the fact 
	\[
	\det \left.\frac{\partial R}{\partial z}\right |_{\zsm}  = \det \left .\begin{bmatrix}
		\I & 0 & 0 \\
		\frac{\partial \vs}{\partial \pos} & 		\frac{\partial \vs}{\partial \velo} & 		\frac{\partial \vs}{\partial \p}\\
		0 & 0 & 1
	\end{bmatrix}\right |_{\zsm} = \det  	\frac{\partial \vs}{\partial \velo}  = -1,
	\]
	where $\vs$ is the velocity component of $R(\pos, \velo,  \p)$, shows that $|\det M| = 1$.
	
	To show $|\det D\Phi_\Tt| = 1$, it now remains to show $ w^\transpose M^{-1} u = -2$. 
	We start by observing that
	\begin{equation}\label{eq:wMu}
		\begin{split}
			w^\transpose M^{-1} u &= \frac{\partial \ts}{\partial z_0} \left( \left.\frac{\partial \phi}{\partial z}\right|_{\Tt-\ts,  \zsp} 
			\left.\frac{\partial R}{\partial z}\right |_{\zsm} 
			\left.\frac{\partial \phi}{\partial z}\right |_{\ts, z_0} \right)^{-1} \\
			&\quad \times \left( \left.-\frac{\partial \phi}{\partial t}\right|_{\Tt-\ts,  \zsp}  + \left.\frac{\partial \phi}{\partial z}\right|_{\Tt-\ts,  \zsp} 	
			\left.\frac{\partial R}{\partial z} \right |_{\zsm} 
			\left.\frac{\partial \phi}{\partial t} \right |_{\ts, z_0} \right)\\
			&= \frac{\partial \ts}{\partial z_0}  \left( 
			\left.\frac{\partial \phi}{\partial z}\right |_{\ts, z_0}^{-1} 
			\left.\frac{\partial R}{\partial z}\right |_{\zsm}^{-1} 
			\left.\frac{\partial \phi}{\partial z}\right|_{\Tt-\ts,  \zsp}^{-1} \right)   \\
			&\quad \times \left( \left.-\frac{\partial \phi}{\partial t}\right|_{\Tt-\ts,  \zsp} + \left.\frac{\partial \phi}{\partial z}\right|_{\Tt-\ts,  \zsp} 	
			\left.\frac{\partial R}{\partial z} \right |_{\zsm} 
			\left.\frac{\partial \phi}{\partial t} \right |_{\ts, z_0} \right)\\
			&= \frac{\partial \ts}{\partial z_0} \left.\frac{\partial \phi}{\partial z}\right |_{\ts, z_0}^{-1} 
			\left( \left.\frac{\partial \phi}{\partial t}  \right |_{\ts, z_0} - \left.\frac{\partial R}{\partial z}\right |_{\zsm}^{-1} 
			\left.\frac{\partial \phi}{\partial z}\right|_{\Tt-\ts,  \zsp}^{-1}  \left.\frac{\partial \phi}{\partial t}\right|_{\Tt-\ts,  \zsp}  \right),
		\end{split}
	\end{equation}
	and simplifying the last expression through the following identity:
	\begin{equation}\label{eq:intermediate_id}
		\left.\frac{\partial \ts}{\partial z_0} \frac{\partial \phi}{\partial z}\right |_{\ts, z_0}^{-1} = 	\begin{bmatrix}
			\0
			& \frac{\partial \ts}{\partial \p_0} 
		\end{bmatrix}.
	\end{equation}
	To establish \eqref{eq:intermediate_id}, we let $A^*_-$ denote $\frac{\partial(\pos^*_-,\velo^*_-)}{\partial(\pos_0, \velo_0)}$ and apply the block matrix inversion formula to obtain:
	\begin{align*}
		\frac{\partial \ts}{\partial z_0} \left.\frac{\partial \phi}{\partial z}\right |_{\ts, z_0}^{-1} &= \begin{bmatrix}	
			\frac{\partial \ts}{\partial \pos_0}	& \frac{\partial \ts}{\partial \velo_0} & \frac{\partial \ts}{\partial \p_0} 
		\end{bmatrix}
		\begin{bmatrix}
			\frac{\partial \pos^*_-}{\partial \pos_0} & \frac{\partial \pos^*_-}{\partial \velo_0} & \frac{\partial \pos^*_-}{\partial \p_0} \\
			\frac{\partial \velo^*_-}{\partial \pos_0} & \frac{\partial \velo^*_-}{\partial \velo_0} & \frac{\partial \velo^*_-}{\partial \p_0} \\
			\frac{\partial \p^*_-}{\partial \pos_0} & \frac{\partial \p^*_-}{\partial \velo_0} & \frac{\partial \p^*_-}{\partial \p_0} \\
		\end{bmatrix}^{-1} & \\
		&=     \begin{bmatrix}	
			\frac{\partial \ts}{\partial \pos_0}	& \frac{\partial \ts}{\partial \velo_0} & \frac{\partial \ts}{\partial \p_0} 
		\end{bmatrix}
		\begin{bmatrix}
			A^*_- & \0  \\
			\begin{bmatrix}\frac{\partial \p^*_-}{\partial \pos_0} & \frac{\partial \p^*_-}{\partial \velo_0}\end{bmatrix} & 1 \\
		\end{bmatrix}^{-1}\\
		&= \begin{bmatrix}	
			\frac{\partial \ts}{\partial \pos_0}	& \frac{\partial \ts}{\partial \velo_0} & \frac{\partial \ts}{\partial \p_0} 
		\end{bmatrix}
		\begin{bmatrix}
			A^{*^{-1}}_- & \0  \\
			-\begin{bmatrix}\frac{\partial \p^*_-}{\partial \pos_0} & \frac{\partial \p^*_-}{\partial \velo_0}\end{bmatrix}	A^{*^{-1}}_- & 1 \\
		\end{bmatrix}\\
		&=
		\begin{bmatrix}
			\left(
			\begin{bmatrix}	
				\frac{\partial \ts}{\partial \pos_0}	& \frac{\partial \ts}{\partial \velo_0} 
			\end{bmatrix}   - 
			\frac{\partial \ts}{\partial \p_0} 
			\begin{bmatrix}\frac{\partial \p^*_-}{\partial \pos_0} & \frac{\partial \p^*_-}{\partial \velo_0}
			\end{bmatrix}\right )	A^{*^{-1}}_-
			& \frac{\partial \ts}{\partial \p_0} 
		\end{bmatrix}\\
		&=
		\left.
		\begin{bmatrix}
			\begin{bmatrix}	
				\frac{\partial \ts}{\partial \pos_0}- \frac{\partial \ts}{\partial \p_0}\frac{\partial \p^*_-}{\partial \pos_0} & \frac{\partial \ts}{\partial \velo_0}  - \frac{\partial \ts}{\partial \p_0}\frac{\partial \p^*_-}{\partial \velo_0}
			\end{bmatrix}  	A^{*^{-1}}_-
			& \frac{\partial \ts}{\partial \p_0} 
		\end{bmatrix}\right |_{\ts, z_0}.
	\end{align*}
	Now consider inertia component of the dynamics $\p(t, z)$, so we have the identity 
	\begin{equation}\label{eq:inertia_id}
		\p\{\ts(z_0), z_0\} =0. 
	\end{equation}
	Differentiating with respect to $\pos_0$, we have
	\[
	\frac{\partial \p}{\partial t}\frac{\partial \ts}{\partial \pos_0} + \frac{\partial \p}{\partial \pos_0} = 0, 
	\]
	which after some rearrangement gives
	\[
	\frac{\partial \ts}{\partial \pos_0} -\frac{\partial \ts}{\partial \p_0} \frac{\partial \p}{\partial \pos_0} = 0 .
	\]
	An identical calculation when differentiating with respect to $\velo_0$ yields 
	\[\frac{\partial \ts}{\partial \velo_0}  - \frac{\partial \ts}{\partial \p_0}\frac{\partial \p}{\partial \velo_0} =0,\]
	and thus 
	\begin{align*}
		\left.\frac{\partial \ts}{\partial z_0} \frac{\partial \phi}{\partial z}\right |_{\ts, z_0}^{-1} &=
		\begin{bmatrix}
			\begin{bmatrix}	
				\0 & \0
			\end{bmatrix}  A^{-1}
			& \frac{\partial \ts}{\partial \p_0} 
		\end{bmatrix}
		=
		\begin{bmatrix}
			\0
			& \frac{\partial \ts}{\partial \p_0} 
		\end{bmatrix}.
	\end{align*}
	Having established the identity \eqref{eq:intermediate_id}, we substitute it back into the expression \eqref{eq:wMu} for $w^\transpose M^{-1} u$ to obtain
	\begin{align*}
		w^\transpose M^{-1} u &= 	 \frac{\partial \ts}{\partial z_0} \left.\frac{\partial \phi}{\partial z}\right |_{\ts, z_0}^{-1} 
		\left( \left.\frac{\partial \phi}{\partial t} \right |_{\ts, z_0}  - \left.\frac{\partial R}{\partial z}\right |_{\zsm}^{-1} 
		\left.\frac{\partial \phi}{\partial z}\right|_{\Tt-\ts,  \zsp}^{-1}  \left.\frac{\partial \phi}{\partial t}\right|_{\Tt-\ts,  \zsp}  \right)\\
		&= 	 
		\begin{bmatrix}
			\0
			& \frac{\partial \ts}{\partial \p_0} 
		\end{bmatrix} 
		\left.\frac{\partial \phi}{\partial t} \right |_{\ts, z_0} 
		-
		\begin{bmatrix}
			\0
			& \frac{\partial \ts}{\partial \p_0} 
		\end{bmatrix} \left.\frac{\partial R}{\partial z}\right |_{\zsm}^{-1} 
		\left.\frac{\partial \phi}{\partial z}\right|_{\Tt-\ts,  \zsp}^{-1}  \left.\frac{\partial \phi}{\partial t}\right|_{\Tt-\ts,  \zsp}. 
	\end{align*}
	
	We now complete the proof by showing that, in the last expression, the first term equals -1 and the second term 1.
	For the first term, we differentiate Equation \eqref{eq:inertia_id} with respect to $\p_0$ to obtain to identity
	\begin{equation}\label{eq:inertia_id2}
		-\frac{\partial \ts}{\partial \p_0}  = \left(\frac{\partial \p}{\partial t}\right)^{-1},
	\end{equation}
	and hence
	\[
	\begin{bmatrix}
		\0
		& \frac{\partial \ts}{\partial \p_0} 
	\end{bmatrix} 
	\left.\frac{\partial \phi}{\partial t} \right |_{\ts, z_0} =  \frac{\partial \ts}{\partial \p_0} \frac{\partial \p^*_-}{\partial t} = -1.
	\]
	For the second term, we first observe that 
	\[
	\begin{bmatrix}
		\0
		& \frac{\partial \ts}{\partial \p_0} 
	\end{bmatrix}
	\left.\frac{\partial R}{\partial z}\right |_{\zsm}^{-1}
	=\begin{bmatrix}
		\0
		& \frac{\partial \ts}{\partial \p_0} 
	\end{bmatrix}
	\ \text{ because } \
	\left.\frac{\partial R}{\partial z}\right |_{\zsm}^{-1} =   \begin{bmatrix}
		\cdot &  \cdot & \cdot \\
		\cdot & \cdot & \cdot \\
		\0 &  \0 & 1 \\
	\end{bmatrix},
	\]
	where the top two rows of $\left.\frac{\partial R}{\partial z}\right |_{\zsm}^{-1}$ get multiplied by 0. 
	So we have
	\begin{align*}
		&\begin{bmatrix}
			\0
			& \frac{\partial \ts}{\partial \p_0} 
		\end{bmatrix}
		\left.\frac{\partial R}{\partial z}\right |_{\zsm}^{-1}
		\left.\frac{\partial \phi}{\partial z}\right|_{\Tt-\ts,  \zsp}^{-1}  \left.\frac{\partial \phi}{\partial t}\right|_{\Tt-\ts,  \zsp}
		=\begin{bmatrix}
			\0
			& \frac{\partial \ts}{\partial \p_0} 
		\end{bmatrix}
		\left.\frac{\partial \phi^{-1}}{\partial z}\right|_{\Tt-\ts, z_\Tt}  \left.\frac{\partial \phi}{\partial t}\right|_{\Tt-\ts,  \zsp}.
	\end{align*}
	And, denoting the inertia component of the inverse map by $\phi_{\p}^{-1}$, we have
	\begin{align}
		\label{eq:intermediate_id_2}
		\begin{bmatrix}
			\0
			& \frac{\partial \ts}{\partial \p_0} 
		\end{bmatrix}
		\left.\frac{\partial \phi^{-1}}{\partial z}\right|_{\Tt-\ts, z_\Tt}  \left.\frac{\partial \phi}{\partial t}\right|_{\Tt-\ts,  \zsp}
		=\frac{\partial \ts}{\partial \p_0} 
		\left.\frac{\partial \phi_{\p}^{-1}}{\partial z}\right|_{\Tt-\ts, z_\Tt }\left.\frac{\partial \phi}{\partial t}\right|_{\Tt-\ts,  \zsp}. 
	\end{align}
	Now consider the identity
	\[
	\phi_{\p}^{-1}(\Tt-\ts, z_\Tt) = \phi_{\p}^{-1}\{\Tt-\ts + s, \phi(s, z_\Tt)\}, \quad s\geq 0.
	\]
	Since the left hand side is constant, taking derivative with respect to $s$ yields:
	\begin{align*}
		0  &= \frac{\diff}{\diff s} \phi_{\p}^{-1}\{\Tt-\ts + s, \phi(s, z_\Tt)\} \\
		&= \left.\frac{\partial \phi_{\p}^{-1}}{\partial t}\right|_{\Tt-\ts +s, \phi(s, z_\Tt)}  + \left.\frac{\partial \phi_{\p}^{-1}}{\partial z}\right|_{\Tt-\ts+s, \phi(s, z_\Tt)} \left.\frac{\partial \phi}{\partial t} \right|_{s, z_T}.
	\end{align*}
	Since this holds for all $s\geq0$, we set $s=0$ and rearrange to get the following identity
	\begin{align*}
		-\left.\frac{\partial \phi_{\p}^{-1}}{\partial t}\right|_{\Tt-\ts, z_\Tt}  &= \left.\frac{\partial \phi_{\p}^{-1}}{\partial z}\right|_{\Tt-\ts, z_\Tt} \left.\frac{\partial \phi}{\partial t} \right|_{0, z_T} \\
		&= 
		\left.\frac{\partial \phi_{\p}^{-1}}{\partial z}\right|_{\Tt-\ts, z_\Tt }\left.\frac{\partial \phi}{\partial t}\right|_{\Tt-\ts,  \zsp},
	\end{align*}
	where the last equality follows from the fact that $\phi(0, z_\Tt) = \phi(\Tt-\ts,  \zsp)$.
	Plugging back into Equation~\eqref{eq:intermediate_id_2}, we obtain
	\begin{align*}
		\frac{\partial \ts}{\partial \p_0} 
		\left.\frac{\partial \phi_{\p}^{-1}}{\partial z}\right|_{\Tt-\ts, z_\Tt }\left.\frac{\partial \phi}{\partial t}\right|_{\Tt-\ts,  \zsp}  
		&=-\frac{\partial \ts}{\partial \p_0} \left.\frac{\partial \phi_{\p}^{-1}}{\partial t}\right|_{\Tt-\ts, z_\Tt}.
	\end{align*}
	Finally, we obtain using Equation \eqref{eq:inertia_id2} 
	\begin{align*}
		-\frac{\partial \ts}{\partial \p_0} \left.\frac{\partial \phi_{\p}^{-1}}{\partial t}\right|_{\Tt-\ts, z_\Tt} 
		&= 
		\left(\left.\frac{\partial \p}{\partial t}\right|_{\ts, z_0}\right)^{-1}
		\left.\frac{\partial \phi_{\p}^{-1}}{\partial t}\right|_{\Tt-\ts, z_\Tt}\\
		&=
		-\frac{1}{\vsm^\transpose \nabla \Ud(\poss)}
		R_\poss(\vsm)^\transpose \nabla \Ud(\poss)\\
		&=1.
	\end{align*}
	We have thus shown $w^\transpose M^{-1} u = -2$ and $|\det D\Phi_\Tt| = |(1+ w^\transpose M^{-1} u)\det M| = 1$. 
	\endgroup
\end{proof}

\section{Proof of Theorem \ref{main-thm:strong}}\label{supp-pf:strong}

\begin{proof}
	To construct a coupled sequence of \bhd{} and \PDMP{}s, let $\pos_0$ and $\velo_0$ denote the initial position and velocity, and let $\{\e_n\}_{n=0, 1, 2, \dots}$ and $\{w_{t,n}\}_{t>0, n=0,1,2,\dots}$ be a collection of independent exponential random variables with unit rate.
	We will couple the two processes through $\e_n$, which will be used both for the inertia refreshments of the \bhd{} and for the simulation of the inhomogeneous Poisson process for the \PDMP{}. The random variables $w_{t,n}$ will be used for additional bounces by the \PDMP{}.

	First we construct the \bhd{} with periodic inertia refreshment at $\Delta t$ intervals. 
	We denote its position, velocity, and inertia components at time $t$ as $\{\pos^\superH_{\Delta t}(t), \velo^\superH_{\Delta t}(t), \p^\superH_{\Delta t}(t)\}$.
	Set $\pos^\superH_{\Delta t}(0) = \pos_0$, $\velo^\superH_{\Delta t}(0) = \velo_0$, and $\p^\superH_{\Delta t}(0)=\e_0$.
	From time $0$ to $\Delta t$, the position and velocity evolve according to the surrogate dynamics and the inertia evolves according to Equation \eqref{main-eq:generaldynamics} in the main text.
	Then at time $\Delta t$, we refresh the inertia by setting $\p^\superH_{\Delta t}(\Delta t) = \e_1$ while maintaining the position and velocity. 
	The dynamics then continues with this new inertia value during $[\Delta t, 2\Delta t]$.
	We repeat this process for time intervals $[n\Delta t, (n+1)\Delta t]$ for $n=2, 3,\dots$ until a total travel time of $\Tt$ is achieved.
	
	Now we construct the \PDMP{} based on the same surrogate dynamics.
	Our construction follows the framework of \citet{vanetti2017piecewise}, but carefully couples the \PDMP{} to its bouncy Hamiltonian counterpart above.
	We denote the position and velocity components as $\{\pos^\superP_{\Delta t}(t), \velo^\superP_{\Delta t}(t)\}$.
	Again, we set  $\pos^\superP_{\Delta t}(0) = \pos_0$ and $\velo^\superP_{\Delta t}(0) = \velo_0$.
	For $t \in [0, \Delta t]$, the process travels according to the surrogate Hamiltonian dynamics until the earlier of $\Delta t$ or the first event at time $\tau^{(1)}_0$, determined by the same exponential variable $\e_0$ used by the \bhd{} with inertia refreshment:
	\[
	\tau^{(1)}_0 = \inf_{t>0} \left \{\e_0 = \int_0^t [\velo^\superP_{\Delta t}(s)^\transpose \nabla U\{\pos^\superP_{\Delta t}(s)\}]^+ \diff s \right\}.
	\]
	\noindent If $\tau^{(1)}_0 \leq \Delta t$, then at time $\tau^{(1)}_0$ the velocity is reflected:
	\[
	\velo^\superP_{\Delta t}(\tau^{(1)}_0) = R_{\pos^{\scalebox{0.8}{\superP}}_{\Delta t}\left(\tau^{(1)}_0\right)}\left\{\lim_{t\uparrow \tau^{(1)}} \velo^\superP_{\Delta t}(t)\right\}.
	\]	
	\noindent The process then continues until time $\Delta t$ or the next event at time $\tau^{(2)}_0 = \tau^{(1)}_0 + t^{(2)}_0$, determined now by $w_{t,n}$:
	\[
	t^{(2)}_0 = \inf_{t>0} \left (w_{\tau^{(1)}_0, 0} = \int_{\tau^{(1)}_0}^{\tau^{(1)}_0+t} [\velo^\superP_{\Delta t}(s)^\transpose \nabla U\{\pos^\superP_{\Delta t}(s)\}]^+ \diff s \right).
	\]
	\noindent If $\tau^{(2)}_0 \leq \Delta t$, another reflection occurs. 
	Additional bounces are computed until the next bounce time $\tau^{(k)}_0$ is greater than $\Delta t$, at which point no further bounces occur on $[0, \Delta t]$.
	
	The above construction connects the \PDMP{}'s evolution on $[0, \Delta t]$ to its bouncy Hamiltonian counterpart's through $\e_0$ in determining the first bounce.
	The subsequent bounces of the \PDMP{} on $[0, \Delta t]$ are dictated by the $w_{t,0}$ and are no longer connected, but become increasingly irrelevant as $\Delta t \to 0$.
	In order to keep the processes connected on future intervals  $[n\Delta t, (n+1)\Delta t]$, we define the first event time within the interval through the shared $\e_n$ by setting the event time as $\tau^{(1)}_n = \min \{n\Delta t + t^{(1)}_n, (n+1) \Delta t\}$, where
	\[
	t^{(1)}_n = \inf_{t>0} \left (\p^\superH(n\Delta t) =  \e_n = \int_{n\Delta t}^{n\Delta t +t} [\velo^\superP_{\Delta t}(s)^\transpose \nabla U\{\pos^\superP_{\Delta t}(s)\}]^+ \diff s \right).
	\]
	
	\noindent The \PDMP{}'s evolution on the remainder of  $[n\Delta t, (n+1)\Delta t]$ is defined in terms of the $w_{t,n}$'s as before.
	
	We now show that these two processes' paths coincide almost surely as $\Delta t \to 0$.
	Observe that
	\begin{align*}
		&\pr\{(\pos_{\Delta t}^\superH, \velo_{\Delta t}^\superH)\equiv (\pos_{\Delta t}^\superP, \velo_{\Delta t}^\superP) \text{ on } [0, T]\}\\
		&\leq \pr\{(\pos_{\Delta t}^\superH, \velo_{\Delta t}^\superH)\equiv (\pos_{\Delta t}^\superP, \velo_{\Delta t}^\superP) \text{ on } [0, \lceil T/\Delta t \rceil \Delta t]\}\\
		&= \prod_{n=0}^{\lceil T/\Delta t \rceil}\pr\{(\pos_{\Delta t}^\superH, \velo_{\Delta t}^\superH)\equiv (\pos_{\Delta t}^\superP, \velo_{\Delta t}^\superP) \text{ on } [n\Delta t, (n+1)\Delta t]\mid(\pos_{\Delta t}^\superH, \velo_{\Delta t}^\superH)\equiv (\pos_{\Delta t}^\superP, \velo_{\Delta t}^\superP) \text{ on } [0,n\Delta t] \}\\
		&= \prod_{n=0}^{\lceil T/\Delta t \rceil}\pr\{(\pos_{\Delta t}^\superH, \velo_{\Delta t}^\superH)\equiv (\pos_{\Delta t}^\superP, \velo_{\Delta t}^\superP) \text{ on } [n\Delta t, (n+1)\Delta t]\mid(\pos_{\Delta t}^\superH, \velo_{\Delta t}^\superH)(n\Delta t) = (\pos_{\Delta t}^\superP, \velo_{\Delta t}^\superP)(n\Delta t) \}.
	\end{align*}
	where the last equality follows from the processes' Markov properties. Now we apply Lemma \ref{lemma:1} to the last expression:
	\begin{align*}
		&\prod_{n=0}^{\lceil T/\Delta t \rceil}\pr\{(\pos_{\Delta t}^\superH, \velo_{\Delta t}^\superH)\equiv (\pos_{\Delta t}^\superP, \velo_{\Delta t}^\superP) \text{ on } [n\Delta t, (n+1)\Delta t]\mid(\pos_{\Delta t}^\superH, \velo_{\Delta t}^\superH)(n\Delta t) = (\pos_{\Delta t}^\superP, \velo_{\Delta t}^\superP)(n\Delta t) \} \\
		&\geq (1-C_T(\pos_0, \velo_0)\Delta t^2)^{ T/\Delta t  +1}\\
		&\geq 1- C_T(\pos_0, \velo_0)(T+\Delta t)\Delta t. 
	\end{align*}
	
	\noindent The last lower bound on the probability tends to 1 as $\Delta t \to 0$, completing the proof.
\end{proof}

\begin{lemma}\label{lemma:1}
	Let $D$ be a finite constant satisfying $\lVert \nabla \Us(\pos) \rVert < D\lVert \pos \rVert + D$ as in Lemma~\ref{lemma:2} and define 
	\[
	M_{s}(\pos, \velo) = \sqrt{\exp\{(D+1) s\}\{(\lVert \pos \rVert+1)^2 + \lVert \velo \rVert^2\}} < \infty
	\]
	so that the solution of bouncy Hamiltonian dynamics from initial condition $(\pos, \velo)$ is guaranteed by the lemma
	to stay within $B_{M_{s}(\pos, \velo) }$, a ball of radius $M_{s}(\pos, \velo)$ centered at the origin.
	Also define 
	\[\kappa_{s}(\pos, \velo) = \max \{\sigma\{\nabla^2 \Ud(\pos')\}: \pos' \in B_{M_{s}(\pos, \velo) } \}   \]
	\[L_{s}(\pos, \velo) = \max \{\lVert \nabla \Ud(\pos') \rVert: \pos' \in B_{M_{s}(\pos, \velo) } \} \]
	
	\noindent where $\sigma(A)$ is the largest singular value of $A$.
	Then,
	\begin{equation}\label{eq:lem1} 
		\begin{split}
			&P\left \{
			\begin{aligned}
				&(\pos_{\Delta t}^\superH, \velo_{\Delta t}^\superH)\equiv (\pos_{\Delta t}^\superP, \velo_{\Delta t}^\superP) \\
				&\text{ on } [n\Delta t, (n+1)\Delta t]
			\end{aligned}\ \middle|\ (\pos_{\Delta t}^\superH, \velo_{\Delta t}^\superH)(n\Delta t) = (\pos_{\Delta t}^\superP, \velo_{\Delta t}^\superP)(n\Delta t) = (\pos_n,\velo_n) \right \}\\
			&\geq 1- C_{\Delta t}(\pos_n,\velo_n)\Delta t^2\\
		\end{split}
	\end{equation}
	where	
	\begin{align*}
		&C_{s}(\pos, \velo) = \max\{ 3 \gamma_{s}(\pos, \velo),  M_{s}(\pos, \velo) L_{s}(\pos, \velo) \} \\
		&\hspace*{2em}\text{for } \ \gamma_{s}(\pos, \velo) = M_{s}(\pos, \velo)^2 \kappa_{s}(\pos, \velo) +  D L_{s}(\pos, \velo)   \{M_{s}(\pos, \velo) + 1\}.
	\end{align*}
	Also, since Lemma~\ref{lemma:2} guarantees
	$M_s(
	\pos_n,\velo_n
	) \leq M_{s + n\Delta t}(\pos_0, \velo_0)$ 
	for any $n$ and $s \geq 0$, it immediately follows that $C_{\Delta t}(\pos_n,\velo_n) \leq C_{\Tt}(\pos_0, \velo_0)$ and that 
	\begin{equation*}
		\begin{split}
			&P\left\{
			\begin{aligned}
				& (\pos_{\Delta t}^\superH, \velo_{\Delta t}^\superH)\equiv (\pos_{\Delta t}^\superP, \velo_{\Delta t}^\superP)\\
				& \text{ on } [n\Delta t, (n+1)\Delta t]
			\end{aligned} \ \middle| \ (\pos_{\Delta t}^\superH, \velo_{\Delta t}^\superH)(n\Delta t) = (\pos_{\Delta t}^\superP, \velo_{\Delta t}^\superP)(n\Delta t)  = (\pos_n, \velo_n)\right\}\\
			&\geq 1- C_{\Tt}(\pos_0, \velo_0) \Delta t^2.
		\end{split}
	\end{equation*} 
\end{lemma}

\begin{proof}
	We prove the result for $n=0$ and the interval $[0, \Delta t]$ with $(\pos_{\Delta t}^\superH, \velo_{\Delta t}^\superH)(0) = (\pos_0, \velo_0)$; the proof is essentially identical for other $n$ since the probability in Equation \eqref{eq:lem1} conditions on the two processes being in identical states at the start of each interval.
	We start by noting that the following inequality holds for $s \leq \Delta t$ regarding the position-velocity component $(\pos_s, \velo_s)$ of the surrogate dynamics from initial position-velocity condition $(\pos_0, \velo_0)$:
	\begin{align*}
		&|\velo_s^\transpose\nabla \Ud(\pos_s) - \velo_0^\transpose \nabla \Ud(\pos_0)| \\
		&= | \velo_s^\transpose\nabla \Ud(\pos_s) -  \velo_s^\transpose\nabla \Ud(\pos_0) +  \velo_s^\transpose\nabla \Ud(\pos_0) - \velo_0^\transpose \nabla \Ud(\pos_0)|\\
		&\leq | \velo_s^\transpose\{\nabla \Ud(\pos_s) -  \nabla \Ud(\pos_0)\} | + |  (\velo_s- \velo_0)^\transpose\nabla \Ud(\pos_0)| \\ 
		&\leq \lVert \velo_s \rVert  \lVert \nabla \Ud(\pos_s) -  \nabla \Ud(\pos_0) \rVert  +  \lVert \velo_s- \velo_0 \rVert L_{\Delta t}(\pos_0, \velo_0) \\ 
		&= \lVert \velo_s \rVert  \left \lVert \int_0^s \velo_u^\transpose \nabla^2 \Ud(\pos_u) \diff u \right \rVert  + L_{\Delta t}(\pos_0, \velo_0) \left\lVert \int_0^s -\nabla \Us(\pos_u)\diff u \right\rVert   \\
		&\leq \lVert \velo_s \rVert  \int_0^s  \lVert \velo_u^\transpose \nabla^2 \Ud(\pos_u) \rVert \diff s  + L_{\Delta t}(\pos_0, \velo_0) \int_0^s \lVert -\nabla \Us(\pos_u) \rVert  \diff u \\
		&\leq \lVert \velo_s \rVert  \int_0^s  \lVert \velo_u \rVert \kappa_{\Delta t}(\pos_0, \velo_0)  \diff u + L_{\Delta t}(\pos_0, \velo_0) \int_0^s (D \lVert \pos_u \rVert   +D )\diff u   \\
		&\leq    s M_{\Delta t}(\pos_0, \velo_0)^2 \kappa_{\Delta t}(\pos, \velo) + s L_{\Delta t}(\pos_0, \velo_0)  D \{M_{\Delta t}(\pos_0, \velo_0) + 1\}    \\
		&=  s \gamma_{\Delta t}(\pos_0, \velo_0).
	\end{align*}
	We have thus shown
	\begin{equation}\label{eq:bound1}
		|\velo_s^\transpose\nabla \Ud(\pos_s) - \velo_0^\transpose \nabla \Ud(\pos_0)| \leq s\gamma_{\Delta t}(\pos_0, \velo_0) \leq \Delta t \gamma_{\Delta t}(\pos_0, \velo_0) \ \textrm{for} \ s \leq \Delta t.
	\end{equation}
	
	Now we upper bound the probability that the chains diverge. 
	We first consider the case 
	\[
	|\velo_0^\transpose \nabla \Ud(\pos_0)| \leq 2\Delta t \gamma_{\Delta t}(\pos_0, \velo_0).
	\]
	In this case, Equation \eqref{eq:bound1} implies $|\velo_s^\transpose\nabla \Ud(\pos_s)| \leq 3 \Delta t \gamma_{\Delta t}(\pos_0, \velo_0)$ and we have the following bound on the probability that the \PDMP{} bounces at least once:
	\begin{align*}
		\pr \left[\e_0 \leq  \int_0^{\Delta t} \{\velo_s^\transpose\nabla \Ud(\pos_s)\}^+\diff s\right]  &\leq \pr\left\{\e_0 \leq  \int_0^{\Delta t} |\velo_s^\transpose\nabla \Ud(\pos_s)|\diff s \right\} \\
		&\leq \pr\left\{\e_0 \leq  \int_0^{\Delta t}  3 \Delta t \gamma_{\Delta t}(\pos_0, \velo_0) \diff s\right\}\\
		&= \pr\{\e_0 \leq 3 \Delta t^2 \gamma_{\Delta t}(\pos_0, \velo_0) \} \\
		& = 1 - \exp\{-3 \Delta t^2\gamma_{\Delta t}(\pos_0, \velo_0)\}\\
		&\leq 3 \Delta t^2 \gamma_{\Delta t}(\pos_0, \velo_0). \yesnumber \label{eq:bpsonebounce}
	\end{align*}
	The last quantity also bounds the probability of the two processes diverging since this event requires at least once bounce and hence is a subset.
	
	We next turn to the case
	\[
	|\velo_0^\transpose \nabla \Ud(\pos_0)| > 2\Delta t \gamma_{\Delta t}(\pos_0, \velo_0),
	\]
	where we have either $\velo_0^\transpose \nabla \Ud(\pos_0)< -2 \Delta t \gamma_{\Delta t}(\pos_0, \velo_0)$ or $\velo_0^\transpose \nabla \Ud(\pos_0) > 2 \Delta t \gamma_{\Delta t}(\pos_0, \velo_0)$.
	If the former holds, Equation \eqref{eq:bound1} implies $\velo_s^\transpose\nabla \Ud(\pos_s)  < -\Delta t \gamma_{\Delta t}(\pos_0, \velo_0) < 0$, and so neither process bounces and the two process stay together.
	If the latter holds,  Equation \eqref{eq:bound1} implies $\velo_s^\transpose\nabla \Ud(\pos_s)  > \Delta t \gamma_{\Delta t}(\pos_0, \velo_0) >  0 $.
	This in particular means that $\{\velo_s^\transpose\nabla \Ud(\pos_s)\}^+ =\velo_s^\transpose\nabla \Ud(\pos_s) $ on $[0, \Delta t]$ and hence neither process bounces or both processes share the first bounce. 
	In the former case, the processes stay together.
	In the latter case, they diverge only if they bounce again.
	To complete the proof, it therefore remains to bound the probability of either process undergoing a second bounce in the case $\velo_0^\transpose \nabla \Ud(\pos_0) > 2\Delta t \gamma_{\Delta t}(\pos_0, \velo_0)$.	
	Additional bounces are impossible for the \bhd{} since an analogous bound as Equation \eqref{eq:bound1} for the post-bounce trajectory implies the inertia is negative for the remainder of the trajectory.
	For the \PDMP{}, the event rate is bounded by $M_{\Delta t}(\pos_0, \velo_0)L_{\Delta t}(\pos_0, \velo_0)$ for $s \in [0,\Delta t]$ and hence its probability of bouncing more than once is bounded by 
	\begin{equation}\label{eq:bpstwobounce}
		\Delta t^2 M_{\Delta t}(\pos_0, \velo_0)  L_{\Delta t}(\pos_0, \velo_0).
	\end{equation}
	
	Combining our bounds from the two cases, we have that the probability of diverging is bounded by $C_{\Delta t}(\pos_0, \velo_0)\Delta t^2$, the larger of Equation \eqref{eq:bpsonebounce} and \eqref{eq:bpstwobounce}.
	This establishes \eqref{eq:lem1} and completes the proof.
\end{proof}

\begin{lemma}\label{lemma:2}
	Assume $\Us(x)$ is continuously differentiable and 
	\[\lim \sup_{\lVert \pos \rVert \to \infty} \frac{\lVert \nabla \Us(\pos) \rVert}{\lVert \pos \rVert} < \infty,\]
	which in particular implies $\lVert \nabla \Us(\pos) \rVert < D\lVert \pos \rVert + D$  for some $D<\infty$.
	Let $\{ (\pos_t, \velo_t) \}_{t \geq 0}$ denote a position-velocity trajectory of either \PDMP{} or \bhd{}.
	Then, regardless of the Poisson bounce locations or of the initial inertia, the trajectory satisfies the following inequality for any $t \geq 0$:
	\[
	\max\{\lVert \pos_t \rVert^2, \lVert \velo_t \rVert^2\} 
	\leq (\lVert \pos_t \rVert+1)^2 + \lVert \velo_t \rVert^2 
	\leq \exp\{(D+1) t\}\{(\lVert \pos_0 \rVert+1)^2 + \lVert \velo_0 \rVert^2\}.
	\]
\end{lemma}

\begin{proof}
	Away from bounce events, the trajectory $\{ (\pos_t, \velo_t) \}_{t \geq 0}$ coincides with a solution of surrogate dynamics.
	Viewed as such, we have the following bound on the quantity $(\lVert \pos_s \rVert+1)^2 + \lVert \velo_s \rVert^2$ for $s \geq 0$:
	\begin{align*}
		\frac{\diff}{\diff s}\left\{(\lVert \pos_s \rVert+1)^2 + \lVert \velo_s \rVert^2\right\}  &= \left\{2(\lVert \pos_s \rVert+1) \frac{\pos_s}{\lVert \pos_s \rVert}\right\}^\transpose \velo_s - 2\velo_s^\transpose \nabla \Us(\pos_s)\\
		&= 2\left\{\pos_s^\transpose \velo_s + \frac{\pos_s^\transpose\velo_s}{\lVert \pos_s \rVert} -  \velo_s^\transpose \nabla \Us(\pos_s)\right\}\\
		&\leq 2\left\{\lVert \pos_s\rVert\lVert \velo_s\rVert + \frac{\lVert\pos_s\rVert\lVert\velo_s\rVert}{\lVert \pos_s \rVert} +  \lVert\velo_s\rVert\lVert \nabla \Us(\pos_s)\rVert\right\}\\
		&= 2\{\lVert \pos_s\rVert\lVert \velo_s\rVert +\lVert\velo_s\rVert +  \lVert\velo_s\rVert\lVert \nabla \Us(\pos_s)\rVert \}\\
		&\leq 2\{ \lVert \pos_s\rVert\lVert \velo_s\rVert +\lVert\velo_s\rVert +  \lVert\velo_s\rVert(D\lVert \pos_s \rVert + D) \}\\
		&= 2(D+1)(\lVert \pos_s\rVert\lVert \velo_s\rVert  +   \lVert\velo_s\rVert)\\
		&\leq (D+1)\{(\lVert \pos_s\rVert +   1)^2 + \lVert \velo_s\rVert^2\}.
	\end{align*}
	By Gronwall's inequality, the above differential inequality implies that a solution of the surrogate dynamics, and hence the trajectory in between bounces, satisfy
	\begin{equation}\label{eq:gron1}
		\begin{split}
			(\lVert \pos_{r+s} \rVert+1)^2 + \lVert \velo_{r+s} \rVert^2 &\leq \exp\{(D+1)s\}\{(\lVert \pos_{r} \rVert+1)^2 + \lVert \velo_{r} \rVert^2\} \\
		\end{split}
	\end{equation}
	for $s,r \geq 0$.
	
	For both the \PDMP{} and \bhd{}, we can partition the interval $[0,t]$ into the subintervals between bounces, and since the bounces do not change the velocity magnitude, we can recursively apply Equation~\eqref{eq:gron1} to each partition to obtain the bound
	\[
	(\lVert \pos_t \rVert +1)^2 +  \lVert \velo_t \rVert^2 \leq \exp\{(D+1) t\}\{(\lVert \pos_0 \rVert+1)^2 + \lVert \velo_0 \rVert^2\},
	\]
	from which the lemma follows directly.
\end{proof}

\section{Is it classical Hamiltonian dynamics in disguise?}
\label{sec:nonhamil}
The prior work of \citet{nishimura2024zigzag} suggests a potential connection between \HMC{} and \PDMP{}s by identifying a novel variant of Hamiltonian dynamics with highly unusual ``bouncy'' behavior.
It is apposite to ask, therefore, whether our \bhd{} can also be described within the existing framework if viewed from a right perspective.
We show this \textit{not} to be the case for the \HBPS{} dynamics and hence for \bhd{} in general. 
Consequently, our bouncy dynamics provides a fundamentally new tool for proposal generation in Bayesian computation.

When considering a univariate target density $\pi(\pos)$, \citet{nishimura2024zigzag}'s Hamiltonian zig-zag coincides with \HBPS{} dynamics in their position-velocity components.
This is analogous to how the \zz{} and \BPS{} coincide in one dimension \citep{bierkens2019zig}.	
In higher dimensions, however, no Hamiltonian dynamics is capable of emulating \HBPS{} dynamics' piecewise linear trajectory in the position space.
In particular, we show that, in contrast to how the Laplace momentum yields Hamiltonian zig-zag, no choice of momentum distribution yields \HBPS{} dynamics:

\begin{theorem}\label{thm:nonhamil}
	Suppose a given target $\pi(\pos)$ in $\mathbb{R}^d$ for $d \geq 3$ is strongly log-concave. 
	Then no choice of smooth momentum distribution $\pi(\mo)$, even allowing for discontinuities in between smooth pieces, yields Hamiltonian dynamics that parallels \HBPS{}'s piecewise linear dynamics in the position component.
\end{theorem}
\noindent
The precise meaning of ``Hamiltonian dynamics'' in the theorem statement warrants clarification. 
Classical smooth Hamiltonian dynamics cannot possibly emulate \HBPS{}'s discontinuous behavior. 
The theorem thus considers a more general class of \textit{discontinuous} Hamiltonian dynamics, in which discontinuous behavior becomes possible through discontinuity in the gradient of the kinetic energy \citep{nishimura2020discontinuous}, and establishes that \HBPS{} dynamics lies outside this even broader class of dynamics. 

When the target is smooth, the discontinuous behavior in Hamiltonian dynamics arises from discontinuities in $K$, which cause instantaneous changes in velocity $\velo_t = \nabla K(p_t)$ when $\mo_t$ crosses the boundary between piecewise smooth components of $K$.
Discontinuous Hamiltonian dynamics in general can be defined only away from a set of Lebesgue measure zero \citep{nishimura2020discontinuous, nishimura2024zigzag}.
In asking for potential existence of an \HBPS{}-like dynamics, therefore, we assume the candidate class of discontinuous Hamiltonian dynamics to be defined away from the union of the following two sets:
\begin{equation}
	\label{eq:problematic_set}
	\{(\pos, \mo): \nabla U(\pos) = 0\} \ \text{ and } \
	\{(\pos,\mo): \mo \in D,  \ \nu(\mo)^\transpose \nabla U(\pos) = 0 \}
\end{equation}
where $D$ is the set of discontinuity boundaries and $\nu(\mo)$ is the vector orthogonal to the boundary.

The above sets parallel those identified by \citet{nishimura2024zigzag} as problematic in guaranteeing well-definedness and uniqueness of Hamiltonian zig-zag, and are thus excluded from the domain of definition.
The first set $\{ \nabla U(\pos) = 0 \}$ is problematic in that it can cause ambiguity in the evolution of momentum, as defined by $\frac{\diff \mo}{\diff t} = \nabla U(\pos)$, as well as in the bounce directions for an \HBPS{}-like Hamiltonian dynamics.
The second set is problematic for the following reason.
Suppose the momentum component of a trajectory $(\pos_t, \mo_t)$ encounters a discontinuity boundary at time $t^*$ so that $\mo_{t^*} \in D$.
As long as $\nu(\mo_{t^*})^\transpose \frac{\diff \mo}{\diff t}(t^*) = \nu(\mo_{t^*})^\transpose \nabla U(\pos_{t^*}) \neq 0$, the trajectory is guaranteed to cross from one side of the boundary to another instantaneously;
this ensures that the velocity $\velo_t = \nabla K(\mo_t)$, and hence the dynamics' behavior, to be uniquely defined by the differential equation except for the instantaneous moment $t = t^*$. 
When $\nu(\mo_{t^*})^\transpose \nabla U(\pos_{t^*}) = 0$, however, there is potential ambiguity in the trajectory's behavior in the next moment. 

We note that, under our strong log-concavity assumption on the target, the first set in \eqref{eq:problematic_set} consists only of a single point and the implicit function theorem implies the second set to consist of manifolds of dimension $2d - 2$. 
As their images under a solution operator of dynamics have measure zero \citep{nishimura2024zigzag}, the exclusion of these problematic sets \eqref{eq:problematic_set} from our definition of the candidate dynamics is reasonable.

\begin{proof}
	Fix an initial condition $(\pos_0, \mo_0)$ outside the measure zero set and away from the discontinuity of $K$. 
	We first observe that, if Hamiltonian dynamics were to generate piecewise linear trajectories, all the trajectories from nearby initial states must bounce in the same fixed direction $\vs$ at their first bounce. 
	This is because, as shown in Lemma \ref{lemma:piecewise_exp}, such Hamiltonian dynamics must have a piecewise exponential $\pi(\mo) \propto \exp\{-K(\mo)\}$ with constant $\nabla K(\mo)$ on each smooth piece.
	And, since the discontinuity set of $K$ is smooth, all the trajectories starting from a sufficiently small neighborhood $B_\delta(\pos_0, \mo_0)$ cross into the same piece with the same constant value of $\nabla K \equiv \vs$ as in the original trajectory $(\pos_0, \mo_0)$.
	
	Now we show that the  above property is incompatible with the \HBPS{}-like bounce mechanism.
	Specifically, we will show that, if the trajectory were to bounce off the hyperplane orthogonal to $\nabla U$, its post-bounce velocity $\tilde \velo^*$ necessarily differs from $\velo^*$ for almost every initial condition in a neighborhood of $(\pos_0, \velo_0)$.
	The demonstration of this incompatibility implies that \HBPS{} dynamics cannot be Hamiltonian and completes the proof.
	
	To establish this, we study the set of initial states within the neighborhood of $(\pos_0, \mo_0)$ from which the first bounce in their trajectories yields $\velo^*$ as the new velocity:
	\[
	\{(\tilde \pos_0, \tilde \mo_0) \in B_\delta(\pos_0, \mo_0): f(\tilde \pos_0, \tilde \mo_0) = \0\} 
	\ \textrm{ for } \
	f(\tilde \pos_0, \tilde \mo_0) = \velo_0- 2 \frac{\velo_0^\transpose \tilde \pos^*}{\lVert  \tilde \pos^* \rVert^2}	\tilde \pos^* - \velo^*.	
	\]
	Once we show that the Jacobian of $f$ evaluated at $(\pos_0, \mo_0)$ has a non-zero rank, it follows from the implicit function theorem that this set is a manifold of dimension less than $2d$ and thus has measure 0.
	\noindent Differentiating $f$ with respect to $\tilde \pos_0$, we obtain
	\begin{align*}
		\left.\frac{\partial f}{\partial \tilde \pos_0} \right|_{\pos_0, \mo_0} &=  \left[\frac{-2}{{\lVert \nabla U(\tilde \pos^*) \rVert^{2}}}\left\{\nabla U(\tilde \pos^*)\velo_0^\transpose - 2 \velo_0^\transpose \nabla U(\tilde \pos^*)\frac{\nabla U(\tilde \pos^*) \nabla U(\tilde \pos^*)^\transpose}{\lVert \nabla U(\tilde \pos^*) \rVert ^2} + \velo_0^\transpose \nabla U(\tilde \pos^*) I\right\}\right.  \\
		&\quad \times \left.\nabla^2U(\tilde \pos^*) \left(I + \velo_0 \left.\frac{\partial t^*}{\partial\tilde \pos_0}\right)\right]\right|_{\tilde \pos_0 = \pos_0, \tilde \mo_0 = \mo_0}.
	\end{align*}
	We see that the right-hand side is a product of three matrices, where the Hessian $\nabla^2U$ has full rank due to strong convexity of $U$ and the other matrices have rank at least $d-1$ being rank-one perturbations of the identity matrix.
	The product is thus at least rank $d-2$ and has a nonzero rank when $d\geq3$, completing the proof.
\end{proof}

\begin{lemma}\label{lemma:piecewise_exp}
	If a Hamiltonian dynamics were to follow a piecewise linear trajectories with constant velocity under a strongly log-concave $\pi(\pos)$ and piecewise smooth $\pi(\mo)$, then $\pi(\mo)$ must be piecewise exponential of the form $\pi(\mo) \propto \exp\{-K(\mo)\}$ for piecewise linear $K(\mo)$.
	
\end{lemma}

\begin{proof}
	Recall the trajectories of Hamiltonian dynamics are governed by the differential Equations \eqref{main-eq:hamildynam} with velocity given by $\velo_t = \nabla K(\mo_t)$. 
	For the trajectories to be piecewise linear, therefore, $\nabla K$ must be piecewise constant along the momentum trajectory $\{\mo_t\}_{t \geq 0}$.
	This trajectory, however, is only a one dimensional path in momentum space and does not immediately imply that $\nabla K$ is piecewise constant globally.
	We will formally show below that we can translate the statement along a trajectory $\{\mo_t\}_{t \geq 0}$ into the global one.
	
	We start the proof by establishing the following property of the dynamics: for any given $\mo_0'$ away from the discontinuities of $K$ and sufficiently small $\epsilon > 0$, the dynamics’ solution operator $\Phi_\epsilon(\pos', \mo_0')$, if viewed as a function $\Phi_{\epsilon, \mo_0' }(\pos')$ only of the position $\pos'$ outputting only the momentum component $\mo_\epsilon'$ of $(\pos_\epsilon', \mo_\epsilon') = \Phi_{\epsilon, \mo_0' }(\pos')$, maps a sufficiently small neighborhood $B(\pos_0')$ of any $\pos_0'$  in the position space bijectively onto a neighborhood $\mathcal{U}\{\mo_\epsilon'(\pos_0')\}$ in the momentum space.
	To establish this, first choose $\epsilon$ small enough to ensure no discontinuity boundaries are crossed.
	The property then follows immediately from the implicit function theorem by observing that the map is given as
	\[
	\mo_\epsilon'(\pos') = \mo_0' + \int_0^{t}\nabla U\{\pos' + s\nabla K(\mo_0')\}\diff{s}, 
	\]
	whose Jacobian is necessarily full rank by the assumption of strong convexity of $U$.
	
	We now use the above property of the solution operator $\Phi_{\epsilon, \mo_0'}: \pos' \to \mo_\epsilon'(\pos' )$ to complete the proof by showing that, for a given $\mo_0$ away from the discontinuities of $K$, we can find a neighborhood of $\mo_0$ on which $\nabla K$ is constant.
	As $\mathcal{U}\{\mo_\epsilon'(\pos_0')\}$ is an image under fixed $\mo_0'$, the piecewise linear nature of solution trajectories guarantee that $\nabla K$ is constant on $\mathcal{U}\{\mo_\epsilon'(\pos_0')\}$.
	Since this statement holds for any $(\pos_0', \mo_0')$ as long as $\mo_0'$ is away from discontinuities of $K$, we can choose $(\pos_0', \mo_0')$ so that $\mo_\epsilon'(\pos_0') = \mo_0$---we can find such $(\pos_0', \mo_0')$, for example, by solving the dynamics backward for duration $\epsilon$ from the initial condition $(\pos_0, \mo_0)$.
	We have thus found a neighborhood $\mathcal{U}\{\mo_\epsilon'(\pos_0')\} = \mathcal{U}(\mo_0)$ on which $\nabla K$ is constant.
\end{proof}

\newcommand{\transKernel}{Q}
\section{Proof of Theorem \ref{main-thm:mh_dominance}}\label{pf:mh_dominance}
Our proof relies on the theory of Peskun-Tierney ordering \citep{peskun1973optimum, tierney1998note}, which states the following.
If transition kernels $\transKernel$ and $\transKernel'$ are both reversible with respect to $\pi(\cdot)$ and satisfy 
\begin{equation}
\label{eq:peskun_tierney_ordering}
\transKernel(\pos \to \pos^*) 
	\leq \transKernel'(\pos \to \pos^*)
	\ \text{ for any } \pos \neq \pos^*,
\end{equation}
then $\transKernel'$ dominates $\transKernel$ in asymptotic variance of the ergodic average $n^{-1} \sum_{i = 1}^{n} f(\pos^{(i)})$ for any function $f$ square integrable with respect to $\pi(\cdot)$.
More precisely, we have
\[
\operatorname{var}(f, \transKernel') \leq \operatorname{var}(f, \transKernel),
\]
where $\operatorname{var}(f, \transKernel)$ denotes the limiting variance of $n^{-1/2} \sum_{i = 1}^{n} f(\pos^{(i)})$ under a homogeneous Markov chain $\pos^{(1)}, \pos^{(2)}, \ldots$ defined by the transition kernel $\transKernel$. 
In particular, Theorem \ref{main-thm:mh_dominance} follows once we establish the ordering \eqref{eq:peskun_tierney_ordering} between the random walk and \HBPS{} samplers' marginal transition kernels in the position variable, which we denote as $\transKernel_\mathrm{RW}$ and $\transKernel_\mathrm{HBPS}$.

\begin{proof}
Our proof will couple the two samplers so that, whenever the random walk sampler achieves an accepted move $\pos \to \pos^*$, the \HBPS{} achieves the identical move. 
This property immediately implies the ordering
$\transKernel_\mathrm{RW}(\pos \to \pos^*) 
	\leq \transKernel_\mathrm{HBPS}(\pos \to \pos^*)$
for any $\pos \neq \pos^*$ and completes the proof.

Fix the current position $\pos$.
Let $\velo \sim N(0,I)$ and $\p \sim \operatorname{Exp}(1)$ denote the auxiliary velocity and inertia variables used to generate the next state of the \HBPS{}.
Couple the random walk Metropolis sampler to the \HBPS{} as follows:
generate the proposal via $\pos^* = \pos + \sigma \velo$, where $\sigma$ is the proposal standard deviation; 
and accept if $\p \geq \Ut(\pos^*) - \Ut (\pos)$, which provides the correct Metropolis acceptance probability.
For the \HBPS{} on a log-concave target, $\p \geq \Ud(\pos^*) - \Ud(\pos)$ implies that no bounces occur during $t \in [0, \sigma]$ and the sampler travels from $\pos$ to $\pos^*$ in its rejection-free manner.
\end{proof}

\bibliography{supplement_ref}